\documentclass[12pt]{article}
\usepackage[dvipsnames]{xcolor}
\usepackage[affil-it]{authblk}
\usepackage{hyperref}
\hypersetup{
 colorlinks,
 citecolor=magenta,
 linkcolor=blue,
 urlcolor=black}
\usepackage[T1]{fontenc}
\usepackage[utf8]{inputenc}
\usepackage[margin=2.5cm]{geometry}
\usepackage{graphicx}
\graphicspath{ {./images/} }
\usepackage{amsmath}
\usepackage{amssymb}
\usepackage{mathtools}
\usepackage{braket}
\usepackage{float}
\usepackage{cancel}
\DeclareMathAlphabet{\mathbbold}{U}{bbold}{m}{n}
\usepackage{parskip}
\usepackage{array}
\usepackage{xfrac}
\numberwithin{equation}{section}
\usepackage{setspace}
\title{{\bfseries{Supersymmetric extensions of Kac–Moody boundary conditions in $\text{AdS}_3$ gravity}}}
\author[a]{Nabamita Banerjee}
\author[a,b]{Vedant Bhutra}
\author[a]{Suvankar Dutta}
\author[a]{Soumava Kundu}
\affil[a]{\small Department of Physics, Indian Institute of Science Education and Research Bhopal, Bhopal Bypass Road, Bhopal - 462066, India}
\affil[b]{\small Department of Theoretical Physics, Tata Institute of Fundamental Research, 1, Homi Bhabha Road, Mumbai - 400005, India}
\date{\small{Email: \texttt{\href{mailto:nabamita@iiserb.ac.in}{nabamita@iiserb.ac.in}}, \texttt{\href{mailto:vedant.bhutra@tifr.res.in}{vedant.bhutra@tifr.res.in}}, \texttt{\href{mailto:suvankar@iiserb.ac.in}{suvankar@iiserb.ac.in}}, \texttt{\href{mailto:soumava21@iiserb.ac.in}{soumava21@iiserb.ac.in}}}\\ \vspace{0.4cm}\today}

\newcommand{\zz}{\mathbb{Z}}
\newcommand{\rr}{\mathbb{R}}
\newcommand{\N}{\mathcal{N}}
\newcommand{\Ell}{\mathcal{L}}
\newcommand{\Q}{\mathcal{Q}}

\newcommand{\slr}{\text{SL}(2,\mathbb{R})}

\newcommand{\osp}{\mathfrak{osp}}
\DeclareMathOperator{\STr}{STr}

\newcommand{\fdel}[2]{\frac{\delta #1}{\delta #2}}

\newcommand{\param}{(t, \theta)}

\begin{document}
\maketitle

\begin{abstract}
We extend the Kac--Moody (KM) boundary conditions of AdS$_3$ gravity by incorporating fermionic fields. For $\mathcal{N}=(1,1)$ AdS$_3$ supergravity, we show that there are two possible ways to implement the fermionic extension. In the first, the extended KM boundary conditions are related to the standard super--Virasoro (VS) boundary conditions through a large gauge transformation realized by the \emph{super--Miura map} between fields and chemical potentials, establishing a supersymmetric generalization of the KM--VS correspondence. In the second, a more general boundary configuration leads to strong constraints on the fermionic chemical potentials, yet offers a much richer asymptotic structure. It provides us a novel realization of the extended Kac–Moody algebra, and a geometric interpretation in terms of folds in the relativistic free-fermion droplet. Finally, we quantize the latter theory by promoting the classical Poisson brackets to (anti-)commutators, construct the corresponding Hilbert space, and show that the resulting spectrum contains only bosonic soft excitations, with no additional fermionic soft modes.
\end{abstract}

\newpage
\tableofcontents

 \section{Introduction and summary}
 
Three-dimensional gravity possesses no local propagating degrees of freedom. Thus, all non-trivial physical information, such as conserved charges, phase-space structure, and asymptotic symmetries, originates solely from the choice of boundary conditions. In their seminal work, Brown and Henneaux \cite{brown1986central} introduced a canonical set of boundary conditions where the temporal ($t$) and angular ($\theta$) components of the gauge fields, interpreted respectively as chemical potentials and dynamical fields, are turned on along specific generators of the gauge group. For constant chemical potentials, these boundary conditions yield an asymptotic symmetry algebra consisting of two copies of the Virasoro algebra, and are regarded as the standard AdS$_3$ boundary conditions. In the following, we will refer to them as the \emph{Virasoro} (VS) boundary conditions.

An alternate class of boundary conditions\footnote{The most general boundary conditions for Einstein gravity on $\text{AdS}_3$ was detailed in \cite{grumiller2016most}. This was generalized to include supersymmetry in \cite{valcarcel2019new}. See also \cite{compere2013new, troessaert2013enhanced, avery2014sl}.} was proposed in \cite{afshar2016soft} and developed in \cite{grumiller2016higher, afshar2017soft}, where the chemical potentials and dynamical fields are turned on along different generators. For constant chemical potentials, this leads to two copies of a $\mathfrak{u}(1)$ Kac--Moody Poisson structure, referred to as the \emph{Kac--Moody} (KM) boundary conditions.\footnote{These are also known as \emph{Heisenberg}, \emph{soft hairy}, or \emph{near horizon} boundary conditions in the literature.} Both the VS and KM setups can be generalized to include field-dependent chemical potentials specified by a functional of the boundary data, called the \emph{boundary Hamiltonian}, which fully determines the boundary dynamics \cite{perez2016boundary, gonzalez2018revisiting, ojeda2019boundary, grumiller2020near,Ojeda:2020bgz, dutta2023higher}. The KM boundary conditions are of particular interest in the context of black-hole physics. Motivated by the soft-hair proposal of Hawking, Perry, and Strominger \cite{hawking2016soft}, Afshar, Grumiller, Setare, and Strominger \cite{afshar2016soft} showed that when the KM boundary conditions are imposed in the near-horizon region of Bañados-Teitelboim-Zanelli (BTZ) black holes, the corresponding charges are integrable in the Regge--Teitelboim sense and act non-trivially on black-hole states without changing $(M,J)$, thereby generating an infinite tower of horizon soft hairs \cite{afshar2016soft, afshar2017near, afshar2017horizon, grumiller2020near}.

An important observation is that the VS and KM boundary conditions are connected by a large gauge transformation \cite{afshar2016soft, grumiller2020interpolating}. This transformation maps the Virasoro configuration to the Kac--Moody one and relates their fields and chemical potentials via the \emph{Miura transformation} \cite{miura1968korteweg}. It preserves the underlying algebraic structure, carrying the Virasoro algebra to the Kac--Moody algebra. Moreover, the Brown-Henneaux Hamiltonian maps, under this transformation, to a Hamiltonian that can be interpreted as the phase-space Hamiltonian of relativistic free fermions on $S^1$ \cite{dutta2023higher}.

Upon supersymmetrisation, the bulk gauge symmetry extends to two copies of $\text{OSp}(1|2,\mathbb{R}) \otimes \text{OSp}(1|2,\mathbb{R})$, thereby enlarging the algebra through the inclusion of fermionic generators. The primary objective of this work is to formulate new boundary conditions for AdS$_3$ supergravity in which fermionic fields and their associated chemical potentials are activated along the fermionic directions.\footnote{In this analysis, we restrict our attention to the minimal supersymmetric extension of the bulk algebra, leaving the investigation of extended cases to future work.} This framework constitutes a fermionic extension of the KM boundary conditions. The preservation of supersymmetry within these configurations depends sensitively on the manner in which the fermionic boundary fields are switched on, thereby giving rise to two distinct families of boundary conditions.

\begin{itemize}
\item In the first approach, we construct a particular fermionic extension of the KM boundary conditions. As the minimally supersymmetric extensions of the VS boundary conditions are already known \cite{banados1998anti, valcarcel2019new}, one may enlarge the large gauge transformations to include fermionic directions and study how bosonic and fermionic fields map between the VS and KM sides. This turns out to be the \emph{super--Miura transformation} \cite{mathieu1988superconformal, mathieu1988supersymmetric}. The procedure yields integrable charges and a supersymmetric Kac--Moody algebra on the KM side.

\item In the second approach, we consider the most general fermionic extension of the KM setup, turning on independent fermionic fields and chemical potentials while keeping the bosonic sector fixed. The bulk equations of motion then impose stringent constraints on the fermionic chemical potentials. Unlike the first case, the boundary theory in this setup is not supersymmetric, as the boundary field content consists of one bosonic and two fermionic degrees of freedom. Nevertheless, the bulk theory remains fully supersymmetric, with supersymmetry broken only at the level of the boundary conditions.
\end{itemize}

We further analyse the bulk theory corresponding to the second set of boundary conditions. Following \cite{dutta2023higher}, where the bosonic boundary dynamics were interpreted in terms of free fermions on a circle, whose time evolution was shown to correspond to smooth deformations of the Fermi surface, we extend this framework to include the fermionic sector. In this geometric realization of the boundary dynamics, we demonstrate that fermionic excitations, represented by fermion bilinears, manifest as folds in the free-fermion droplet.

Finally, we proceed to quantize the system by promoting the classical Poisson brackets among the fields to their corresponding (anti-)commutator relations. Within this quantized framework, we identify the ground state and construct the complete Hilbert space. Following \cite{afshar2016soft}, we find that the bosonic soft modes remain soft even in the presence of fermionic contributions to the Hamiltonian, and that no additional fermionic soft excitations emerge in the spectrum.

The paper is organised as follows. Section \ref{sec:Review} provides a brief review of the bosonic KM and VS boundary conditions in AdS$_3$ gravity and their underlying integrable structures. In section \ref{sec:superMiura}, we construct the minimal supersymmetric extension of the KM boundary conditions using the super--Miura transformation, which establishes a direct correspondence with the $\mathcal{N}=1$ super--Virasoro algebra, and reproduces the supersymmetric KdV hierarchy. Section \ref{sec:superKM} introduces an alternate, inequivalent fermionic extension of the Kac--Moody boundary conditions. In contrast to the previous case, this configuration cannot be related to the super--Virasoro structure through any large gauge transformation and instead defines a distinct coupled boson--fermion Poisson structure. We further develop a droplet description of the boundary dynamics for this latter case, providing a geometric interpretation of the phase--space evolution. We then quantize the theory, constructing the oscillator algebra and identifying the spectrum of soft excitations that form the soft--hair sector. Finally, in section \ref{sec:discussion}, we summarize the main results and outline several promising directions for future investigation.

\section{A review of bosonic boundary conditions and their integrable structure}
\label{sec:Review}

In this section, we review the essential aspects of the bosonic \emph{Kac--Moody} and \emph{Virasoro} boundary conditions in $\text{AdS}_3$ gravity. 

In three dimensions, Einstein gravity, with or without a cosmological constant, possesses no local propagating degrees of freedom and is therefore topological, with all non-trivial dynamics residing at the boundary. This feature allows the bulk theory to be reformulated as a three-dimensional Chern--Simons gauge theory, which provides an elegant and manifestly topological description of the dynamics. The Chern--Simons formulation naturally highlights the boundary degrees of freedom and their symmetry algebras, making it an ideal framework for analyzing the boundary dynamics of three-dimensional gravity. In the following, we outline the correspondence between AdS$_3$ gravity and its Chern--Simons formulation, which forms the basis for our discussion of the two sets of boundary conditions considered in this work.

\subsection{3d Einstein gravity as a Chern-Simons theory}
Gravity in the $\text{AdS}_3$ bulk can be recast as a Chern--Simons theory with gauge group $\mathrm{SO}(2,2)$ \cite{achucarro1986chern, witten19882+}. Owing to the algebraic isomorphism  
\begin{equation}
	\mathfrak{so}(2,2) \cong \mathfrak{sl}(2,\mathbb{R}) \oplus \mathfrak{sl}(2,\mathbb{R}),
\end{equation}
the gauge fields naturally split into two chiral sectors, denoted by $A^\pm$. In each sector, the matrix-valued gauge field $A^\pm$ is written in terms of the vielbein $e^a_{\mu}$ and the spin connection $\omega^{ab}_\mu$ as  
\begin{equation}
\label{eq:gaugemetricrel}
A^\pm = \left( \omega^a \pm \frac{e^a}{\ell} \right) T_a,
\end{equation}  
where $\ell$ is the $\text{AdS}$ radius and $T_a$ are the generators of $\slr$. The one-forms $\omega_a$ are related to the spin connection $\omega^{bc}$ via  
\begin{equation}
\omega_a \equiv \frac{1}{2} \epsilon_{abc} \omega^{bc}.
\end{equation}  

With these identifications, the Einstein--Hilbert action can be shown to be equivalent to the difference of two copies of Chern--Simons actions:  
\begin{equation}
\label{eq:CSacn}
    S_{\text{EH}} = S_{\text{CS}}[A^+] - S_{\text{CS}}[A^-],
\end{equation}
 
where each Chern--Simons action takes the form
\begin{equation}
\label{eq:CSacn2}
    S_{\text{CS}}[A^\pm] = \frac{\mathrm{k}}{4\pi} \int \text{Tr} \left[ A^\pm \wedge dA^\pm + \frac{2}{3} (A^\pm)^3 \right] + \mathcal{B}_\infty[A^\pm],
\end{equation}

The level $\mathrm{k}$ is determined by the $\text{AdS}$ radius $\ell$ and the three-dimensional Newton’s constant $G$ through  
\begin{equation}
\label{eq:kGrelation}
\mathrm{k} = \frac{\ell}{4G}.
\end{equation}  

The term $\mathcal{B}_\infty[A^\pm]$ is a boundary contribution ensuring a well-defined variational principle, $\delta S_\text{CS}[A^\pm] = 0$. The trace in \eqref{eq:CSacn2} is taken in the fundamental representation of $\mathfrak{sl}(2, \rr)$. See appendix \ref{appendix:basis} for details on the generators.

Finally, the metric components follow from the gauge fields via the relation  
\begin{equation}\label{eq:metric}
g_{\mu\nu} = \frac{\ell^2}{2} \mathrm{Tr} \big[(A^+_\mu - A^-_\mu) (A^+_\nu - A^-_\nu)\big].
\end{equation}  

The Lorentzian $\text{AdS}_3$ spacetime manifold $\mathcal{M}$ can be described using coordinates $(t, r, \theta)$, where $t$ denotes time, $\theta$ is a compact angular variable and $r$ is the radial coordinate whose limit $r \to \infty$ corresponds to the asymptotic boundary. We express the gauge connections as \cite{coussaert1995asymptotic}
\begin{equation}
\label{eq:Apm}
A^\pm = b_\pm^{-1} \left(d + a^\pm \right) b_\pm ,
\end{equation}  
with $b_\pm$ denoting group elements that depend exclusively on the radial coordinate $r$. The \emph{auxiliary} connections $a^\pm$ are taken to depend only on the boundary coordinates $t$ and $\theta$ and capture the asymptotic nature of the connection. The precise form of $b_\pm$ is left unspecified, as it is not required for satisfying the Chern--Simons equations of motion, which are equivalent to the Einstein equations in the metric formulation. In the following, we will frequently expand
\begin{equation}
	a^\pm \; = \; a^\pm_\theta \, d \theta \: + \: a^\pm_t \, dt.
\end{equation}

\subsection{Kac-Moody boundary conditions}

We consider the connection
\begin{equation}
\label{eq:apmform}
a^\pm(t,\theta) = \big(\xi_\pm(t,\theta) \, dt \pm p_\pm(t,\theta) \, d\theta \big) L_0,
\end{equation}
where $p_\pm (t,\theta)$ denote the dynamical fields of the gravitational sector and $\xi_\pm (t,\theta)$ are the corresponding chemical potentials. 

When the chemical potentials $\xi_\pm$ are field-independent, the resulting boundary conditions lead to a Kac--Moody asymptotic symmetry algebra \cite{afshar2016soft, grumiller2016higher, dutta2023higher}. Allowing the chemical potentials to depend on the dynamical fields renders the boundary theory integrable. We refer to both cases collectively as the KM boundary conditions.

The variation of the action with respect to $A^\pm$ vanishes in the bulk whenever the curvature constraints $F^\pm=0$ hold. Thus, the Chern--Simons equations of motion,  
\begin{equation}
\label{eq:Maxeq}
    F^\pm \equiv dA^\pm + A^\pm \wedge A^\pm = 0,
\end{equation}  
when evaluated using the forms of $A^\pm$ and $a^\pm$, lead to
\begin{equation}
\label{eq:maxeq}
\dot{p}_\pm(t,\theta) = \pm \xi_\pm'(t,\theta),
\end{equation}  
with $\dot{}$ and $'$ denoting derivatives with respect to $t$ and $\theta$, respectively.

\subsubsection{Boundary Hamiltonians and Poisson structure}
\label{section:hamiltonian}

To identify the boundary contribution $\mathcal{B}_\infty^{\pm}$, we impose the requirement that the action remains stationary under arbitrary variations of the gauge fields, namely  
\begin{equation}
	\delta S_\text{CS} = 0.
\end{equation}
Consequently, the variation of the boundary term must take the form \cite{banados1995global, perez2014higher}
\begin{align}
\label{eq:deltaBinfty}
	\delta \mathcal{B}_\infty^\pm \; &= \; - \frac{\mathrm{k}}{2\pi} \int_{\partial \mathcal{M}} dt \wedge d\theta \ \mathrm{Tr}[ a_t^\pm \, \delta a_\theta^\pm ],\\ 
	&= \; \mp \frac{\mathrm{k}}{4 \pi} \int_{\partial \mathcal{M}} dt \wedge d\theta \ \xi_\pm \, \delta p_\pm.
\end{align}

Therefore, to obtain a well-defined boundary term, one must take the variation 
$\delta$ outside the integral. This is possible if the chemical potentials 
$\xi_{\pm}$ can be expressed as functional derivatives of some quantities, denoted by $H^{\pm}$, with respect to $p_{\pm}$, namely
\begin{equation}
\label{eq:xi-H}
\xi_{\pm} \,=\, -\frac{4\pi}{\mathrm{k}}\,\frac{\delta H^{\pm}}{\delta p_{\pm}}.
\end{equation}
In general, $H^\pm$ are just some functionals of the dynamical fields $p_{\pm}$ and their derivatives with respect to $\theta$. Explicitly, one may write
\begin{equation}
	H^\pm \,=\, \int d\theta \, \mathcal{H}^\pm (p_\pm, p'_\pm, p''_\pm, \ldots),
\end{equation} where $\mathcal{H}^{\pm}$ denotes the corresponding functional density. The boundary term thus takes the form  
\begin{equation}
\label{eq:Binfty}
\mathcal{B}_\infty^{\pm} \;=\; \pm \int dt \, d\theta \, \mathcal{H}^{\pm} 
\;=\; \pm \int dt \, H^{\pm}.
\end{equation}

By \eqref{eq:xi-H}, the boundary term and the boundary conditions, and consequently the dynamics of the fields $p_\pm$, are governed by these choice of the functionals $H^{\pm}$. In particular, one finds  
\begin{equation}
\dot{p}_{\pm}(t,\theta) \,=\, \pm \,\xi'_\pm (t, \theta) \,=\, \mp \,\frac{4\pi}{\mathrm{k}}\,\partial_\theta 
\bigg(\frac{\delta H^{\pm}}{\delta p_{\pm}}\bigg).
\end{equation}

Thus, it is natural to give the functions $H^\pm$ the physical interpretation of Hamiltonians in the respective sector. The equations of motion derived above can be used to declare the Poisson structure of the fields $p_{\pm}(t,\theta)$. In particular, requiring that the time 
evolution be generated by the Hamiltonian,  
\begin{equation}
	\dot{p}_{\pm}(t,\theta) \,=\, \{p_{\pm}(t,\theta), \, H^{\pm}\}_{\text{PB}},
\end{equation}
one is led to the following Poisson bracket:  
\begin{equation}
\label{eq:KMPB}
	\{p_{\pm}(t,\theta) , \, p_{\pm}(t,\theta')\}_{\text{PB}} \,=\, \mp \,\frac{4\pi}{\mathrm{k}}\, \partial_\theta \,\delta(\theta - \theta').
\end{equation}

Taking Fourier modes and promoting the Poisson brackets to commutators, one obtains the $\mathfrak{u}(1)$ Kac-Moody algebras of level $\pm \tfrac{1}{2}$.
\begin{equation}
\label{eq:KMalg}
	[p^\pm_n, p^\pm_m] = \pm \frac{\mathrm{k}}{2} n \, \delta_{n+m,0}, \qquad n, m \in \zz.
\end{equation}

\subsubsection{Asymptotic symmetry algebra and conserved charges}
\label{asa_analysis}

The asymptotic symmetries are generated by gauge transformations that preserve the chosen asymptotic form of the gauge fields \eqref{eq:apmform}. Such a transformation takes the form  
\begin{equation}
\label{eq:gtr}
\delta a^{\pm} \,=\, d\lambda^{\pm} + [a^{\pm} ,  \lambda^{\pm}],
\end{equation} 
where $\lambda^{\pm} \equiv \eta_\pm \, L_0 + \eta^1_\pm \, L_1 + \eta^{-1}_\pm \, L_{-1}$ is the gauge transformation parameter. The asymptotic fields $p_{\pm}$ and the chemical potentials $\xi_{\pm}$ transform as  
\begin{equation}
\label{eq:LgaugetransformKM}
\delta p_{\pm} \,=\, \pm \, \eta'_{\pm}, 
\qquad 
\delta \xi_{\pm} \,=\, \dot{\eta}_{\pm},
\end{equation} while the other components of $\lambda^\pm$ generate trivial gauge transformations. Without loss of generality, we can set $\eta^{\pm 1}_\pm  = 0$. The gauge parameters $\eta_\pm$ are arbitrary functions of the time and angular coordinates. The variation of the conserved charges associated with the local gauge symmetry is given by\footnote{The charge may be non-integrable, hence the $\cancel{\delta}$ notation.} \cite{regge1974role, banados1995global}.  
\begin{equation}
\label{conservedcharge}
\cancel{\delta} \mathbb{Q}^\pm [\lambda^\pm] = - \frac{\mathrm{k}}{2 \pi} \oint d \theta \, \mathrm{Tr}[\lambda^\pm \delta a^\pm_\theta].
\end{equation}

Evaluating this expression for our choice of boundary conditions yields
\begin{equation}\label{eq:KMQvar}
\delta \mathbb{Q}^{\pm} [\eta_\pm] = \mp \, \frac{\mathrm{k}}{4\pi} \oint d\theta\,
\eta_\pm \, \delta p_{\pm}.
\end{equation}
In the special case where the chemical potentials $\xi_\pm$ are field-independent, \eqref{eq:LgaugetransformKM} implies that the gauge transformation parameters $\eta_\pm$ are also field-independent.\footnote{In fact, the connection between the chemical potentials $\xi_\pm$ and $\eta_\pm$ runs deeper. Observe that under the choice $\lambda^{\pm 1}_\pm = 0$, the Chern--Simons equations of motion \eqref{eq:Maxeq} and the variation of the time derivative component $a_\theta$ \eqref{eq:gtr} is equivalent up to an arbitrary scaling factor, meaning that $\xi_\pm$ and $\eta_\pm$ satisfy the same equations. In fact, we show below that $\eta_\pm = \xi_\pm$ is a consistent choice for the gauge parameter.} Therefore \eqref{eq:KMQvar} can be trivially functionally integrated and we obtain
\begin{equation}
\label{eq:KMQ}
    \mathbb{Q}^{\pm} [\eta_\pm] = \mp \, \frac{\mathrm{k}}{4\pi} \oint d\theta \, \eta_\pm\, p_{\pm}.
\end{equation}

The Poisson brackets of these global charges follow directly from evaluating the variation of one charge under the symmetry generated by another \cite{regge1974role, blagojevic2001gravitation},   
\begin{equation}
\label{eq:RGformula}
	\delta_{\lambda_2} \mathbb{Q}[\lambda_1] = \{ \mathbb{Q} [\lambda_1], \mathbb{Q} [\lambda_2] \}.
\end{equation}

Using \eqref{eq:LgaugetransformKM} we find that the asymptotic symmetry algebra is given by two copies of the Kac-Moody algebra \eqref{eq:KMalg}. In general, the chemical potentials $\xi_{\pm}$ may depend on the fields $p_{\pm}$ through \eqref{eq:xi-H} and their variations \eqref{eq:LgaugetransformKM} may not vanish. Combining the two, one can express the transformation parameters in terms of $H^\pm$ as  
\begin{equation}
\label{eq:etaeq}
	\dot{\eta}_\pm (t,\theta) \;=\; \mp \, \frac{4\pi}{\mathrm{k}} \, \frac{\delta}{\delta p_{\pm}(t,\theta)} \int d\theta'\, \left(\frac{\delta H^{\pm}}{\delta p_{\pm}} \, \eta'_{\pm}(t,\theta') \right).
\end{equation}

The gauge transformation parameters $\eta_\pm$ are no longer field-independent, and \eqref{eq:etaeq} does not yield a unique solution. For a given choice of $H^\pm$, one can, in principle, introduce many functionals, which we label here by $n$,\footnote{At this point the labelling is arbitrary. In the next subsection we provide an example of a countably-infinite set of such functionals, however one can come up with many more.} namely
\begin{equation}
	H^{\pm}_n \; \equiv \; \int d\theta \,\mathcal{H}^{\pm}_n,
\end{equation}
such that the gauge parameters are generated as  
\begin{equation}
	\eta_{\pm} \;=\; -\,\frac{4\pi}{\mathrm{k}}\,\frac{\delta H^{\pm}_n}{\delta p_{\pm}},
\end{equation}
and they satisfy \eqref{eq:etaeq} on-shell. In this case, the quantities  
\begin{equation}
	\mathbb{Q}^{\pm}_n \; \equiv \; \pm H_n^\pm
\end{equation} 
also represent conserved charges of the theory, corresponding to the boundary conditions specified by the choice of $H^{\pm}$. In fact, one can verify that choosing the gauge parameters as  
\begin{equation}
\label{eq:eta-H}
	\eta_{\pm} \;=\; -\,\frac{4\pi}{\mathrm{k}}\,\frac{\delta H^{\pm}}{\delta p_{\pm}}
\end{equation}
is also a consistent solution to \eqref{eq:etaeq}. Thus, the Hamiltonians themselves qualify as conserved charges of the system. This is consistent with \cite{witten19882+}, which specifies that the Hamiltonian of the system is given by $H \equiv \mathbb{Q}[a^\pm_t]$ which is easily verified on comparing \eqref{eq:xi-H} and \eqref{eq:eta-H}.

 Furthermore, the Poisson brackets among this infinite set of conserved charges vanish on-shell, thereby manifesting the integrable structure of the system.

\subsubsection{A relativistic free fermion droplet description}
\label{sec:rel_free_fermi}
It was shown in \cite{dutta2023higher} that, by choosing the Hamiltonian to be the 
collective field theory Hamiltonian \cite{jevicki1980quantum, jevicki1981collective},\footnote{Up to a constant proportionality factor.} one can construct an infinite sequence of conserved charges expressed as phase space integrals of integer powers of the single-particle Hamiltonian of non-relativistic free fermions on $S^1$ \cite{brezin1978planar, polchinski1991classical}. Thus the time evolution of spin-2 fields $p_\pm$ is captured by the classical dynamics of Fermi surfaces under the appropriate choice of the chemical potentials $\xi_\pm$.

This analysis can be extended to a different integrable structure, where the Hamiltonian is instead taken to be that of relativistic fermions. Consider the single-particle Hamiltonian $\mathfrak{h}$ of a relativistic fermion on $S^1$ parametrized by angle $\theta$,  
\begin{equation}
\mathfrak{h} = p + W(\theta),
\end{equation} where $p$ is the momentum of the fermion and $W$ is an arbitrary potential.
The corresponding phase-space Hamiltonian is  
\begin{equation}
H_0 \;=\; \frac{\mathrm{k}}{4\pi} \int d\theta \, dp \, \big(p + W(\theta)\big)\, \varpi(\theta,p),
\end{equation}  
where $\varpi(\theta, p)$ is the phase-space density, defined as  
\begin{equation}
\varpi(p,\theta) \;=\; \Theta \Big( \big(p_+(t,\theta) - p\big)\,\big(p - p_-(t,\theta)\big) \Big),
\end{equation}  
with $\Theta(x)$ the Heaviside step function. Evaluating this expression gives  
\begin{equation}
H_0 \;=\; \frac{\mathrm{k}}{4\pi} \int d\theta \, \left(\frac{p_+^2}{2} - \frac{p_-^2}{2} + W(\theta)\,(p_+ - p_-) \right).
\end{equation}  

For the relativistic Hamiltonian, one can define an infinite tower of conserved charges as  
\begin{equation}
H_n \;=\; \frac{\mathrm{k}}{4\pi} \int d\theta \, dp \, \big(p + W(\theta)\big)^n \varpi(\theta,p) \; \equiv \; H_n^{+} - H_n^{-}
\end{equation}
such that $\{H_m, \, H_n\}_\text{PB} = 0$.

In absence of $W(\theta)$, one can have conserved charges that explicitly depend on derivatives of $p^\pm$. The first few terms of such a tower of charges is given by
\begin{equation}
\label{towerofcharges}
\begin{split}
		H^\pm_0 &= \frac{\mathrm{k}}{4\pi}\int d\theta \, p_\pm, \\
		H^\pm_1 &= \frac{\mathrm{k}}{4\pi}\int d\theta \, \frac{1}{2} p_\pm^2, \\
		H^\pm_2 &= \frac{\mathrm{k}}{4\pi}\int d\theta \, \frac{1}{3} \big(p_\pm^3 + p_\pm'^{\,2} \big), \\
		H^\pm_3 &= \frac{\mathrm{k}}{4\pi}\int d\theta \, \frac{1}{4} \big(p_\pm^4 + 4p_\pm p_\pm'^{\,2} + \frac{4}{5} p_\pm''^{\, 2} \big). \\
	\end{split}
\end{equation} These are given in terms of Gelfand--Dikii polynomials\footnote{The precise factors depend on the coefficient of the third-derivative term in \eqref{eq:OpD}, which is freely adjustable, and the normalization chosen for the Gelfand--Dikii polynomials.} and form a \emph{family} of boundary conditions labelled by non-negative integers, that corresponds to the \textit{Korteweg-de Vries (KdV) hierarchy} \cite{gonzalez2018revisiting, grumiller2020near}.

One may also use \emph{generalized} Gelfand--Dikii polynomials instead to obtain the \emph{Gardner} or \emph{mixed KdV-mKdV} hierarchy. For details we refer the reader to \cite{ojeda2019boundary}.

\subsubsection{Near horizon symmetries and soft hairy black holes}

So far, we have discussed asymptotic dynamics. However, as the canonical charges \eqref{eq:KMQ} and their algebra \eqref{eq:KMalg} do not depend on the radial coordinate, the same structure arises at any valid radius $r = r_0$. In particular, the metric \eqref{eq:metric} constructed out of the connections $A^\pm$, for constant chemical potentials \cite{afshar2016soft}, generically corresponds to a \emph{black flower} solution to Einstein's equations in 3d with a negative cosmological constant \cite{barnich2016asymptotically}, of which the Bañados-Teitelboim-Zanelli (BTZ) black hole \cite{banados1992black, banados1993geometry} is a special case. Thus, one can take the perspective of a near-horizon observer. It was then shown in \cite{afshar2016soft} that the Hamiltonians of the system commute with the generators of the near horizon symmetry algebra and thus describe a tower of infinite soft modes, which do not contributed to the black hole entropy.

\subsection{Virasoro boundary conditions}

In their seminal work, now recognized as a precursor to the $\text{AdS}/\text{CFT}$ correspondence, Brown and Henneaux \cite{brown1986central} introduced a particular choice of boundary conditions in the highest-weight gauge \cite{henneaux2013chemical,bunster2014generalized},
\begin{equation}
\label{eq:apmthetat}
\begin{split}
	\tilde{a}^\pm_\theta &= L_{\pm 1} - \frac{1}{2} \mathcal{L}_\pm \, L_{\mp 1},\\
	\tilde{a}^\pm_t &= \mu_\pm \, a^\pm_\theta \mp \mu'_\pm \, L_0 +\frac{1}{2} \mu''_\pm \, L_{\mp 1} 
\end{split}
\end{equation}
where $\tilde{a}^\pm = \tilde{a}^\pm_\theta \, d\theta + \tilde{a}^\pm_t \, dt$, and the functions $\mathcal{L}_\pm(t,\theta)$ and $\mu_\pm(t,\theta)$ represent the dynamical fields and the chemical potentials, respectively.\footnote{The original calculation by Brown and Henneaux was performed in the metric formulation, and corresponded to the special case of $\mu_\pm = -1$.} We stress that, at this point, the boundary conditions remain not entirely specified, as a complete definition requires prescribing the exact form of the Lagrange multipliers $\mu_\pm$ at spatial infinity. 

Starting from the field equations \eqref{eq:Maxeq}, one can show that their asymptotic form must satisfy 
\begin{equation}
\label{eq:VSeom}
\dot{\mathcal{L}}_\pm = \pm D_\pm \mu_\pm,
\end{equation}
where the operator $D_\pm$ is given by
\begin{equation}
\label{eq:OpD}
    D_\pm \equiv \left( \partial_\theta \mathcal{L}_\pm \right) + 2 \mathcal{L}_\pm \, \partial_\theta - \partial_\theta^3.
\end{equation}
The rest of the construction goes similar to the Kac-Moody case. In order to construct a boundary term one can show that the chemical potentials need to be made field-dependent through a choice of a boundary Hamiltonian. Under the asymptotic fall-off conditions specified in \eqref{eq:Apm} and \eqref{eq:apmthetat}, the corresponding variation of the boundary term is  
\begin{equation}
\delta \mathcal{B}^\pm_\infty = \mp \frac{\mathrm{k}}{4\pi} \int dt \wedge d\theta \, \mu_\pm \, \delta \mathcal{L}_\pm .
\end{equation}

As the $\mathfrak{sl}(2,\rr)$ connections are independent, the variational principle is consistent and the boundary term is integrable if the Lagrange multipliers are related to a functional $H_\pm$ by  
\begin{equation}
\mu_\pm = - \frac{4\pi}{\mathrm{k}} \, \frac{\delta H^\pm}{\delta \mathcal{L}_\pm},
\end{equation}
where, as before, $H_\pm$ can be any functional of $\mathcal{L}_\pm$ and its derivatives. Integrating the boundary term, one gets \eqref{eq:Binfty}. Therefore, specifying $H_\pm$ at the boundary completely fixes the boundary conditions \cite{perez2016boundary}.

One can then proceed to determine the corresponding asymptotic symmetry structure. Using the fall--off conditions in \eqref{eq:Apm}, this analysis can be performed directly in terms of the auxiliary connections $a^\pm$.

The dynamical equation becomes
\begin{equation}
   \dot{\mathcal{L}}_\pm = \mp \frac{4\pi}{\mathrm{k}}D_\pm \left(\frac{\delta H^\pm}{\delta \mathcal{L}_\pm}\right) . 
\end{equation}
This equation refers to the Poisson structure between the fields,
\begin{equation}
\label{eq:virasoroalgebra}
    \{\mathcal{L}_\pm(t,\theta), \mathcal{L}_\pm(t,\theta')\}_\text{PB} = \mp \,\frac{4\pi}{\mathrm{k}}  D_\pm \, \delta(\theta-\theta').
\end{equation}
A suitable Fourier mode expansion and replacing the Poisson brackets with commutators in \eqref{eq:virasoroalgebra} yields the Virasoro algebra $\textsf{Vir}$,
\begin{equation}
\label{eq:virasoroalgebra_modes}
	[L_n, L_m] = (n-m)L_{n+m} + \frac{c}{12}n^3 \, \delta_{n+m,0}, \qquad m,n \in \zz
\end{equation}in each sector with central charge
\begin{equation}
    c = \frac{3\ell}{2G}.
\end{equation}

We can also derive the conserved charges. As usual, the relevant gauge transformations are given by \eqref{eq:gtr} where we demand that they preserve the functional form of $\tilde{a}^\pm$ given in \eqref{eq:apmthetat}. This condition is satisfied if the gauge parameters take the form
\begin{equation}
	\tilde \lambda^\pm = \tilde{\eta}_\pm \, \tilde{a}^\pm_\theta \,\mp\,\tilde\eta'_{\pm} L_0 +\frac{1}{2} \tilde\eta''_{\pm} L_{\mp 1},
\end{equation}
and the dynamical fields transform according to
\begin{equation}
\label{eq:Lgaugetransform}
\delta \mathcal{L}_\pm = D_\pm \tilde\eta_\pm.
\end{equation}

Requiring invariance of the temporal components $\tilde{a}^\pm_t$ imposes the additional restriction
\begin{equation}
\label{eq:mugaugetransform}
\delta \mu_\pm = \pm \dot{\tilde\eta}_\pm + \tilde\eta_\pm \mu'_\pm - \tilde\eta'_{\pm} \mu_\pm.
\end{equation}

The time evolution of the symmetry parameters is thus determined by $H^\pm$ as
\begin{equation}
\label{eq:epsiloneqn}
\dot{\tilde\eta}_\pm = \mp \frac{4 \pi}{\mathrm{k}} \frac{\delta}{\delta \mathcal{L}_\pm (t, \theta)} 
\int d\theta' \bigg( \frac{\delta H^\pm}{\delta \mathcal{L}_\pm} \, D_\pm \tilde\eta_\pm (t, \theta') \bigg).
\end{equation}

As $\mu_\pm$ can be field-dependent, $\tilde{\eta}_\pm$ generically depend not only on the coordinates $(t,\theta)$ but also on the dynamical fields $\mathcal{L}_\pm$ and their derivatives. In the canonical approach \cite{regge1974role}, the variation of the charge associated with a symmetry parameter $\tilde\eta_\pm$ is given by
\begin{equation}
\label{eq:VSdeltaQ}
\delta \mathbb{Q}^\pm[\tilde\eta_\pm] = \mp\frac{\mathrm{k}}{4\pi} \int d\theta \ \tilde\eta_\pm \, \delta \mathcal{L}_\pm. 
\end{equation}
The condition \eqref{eq:epsiloneqn} guarantees that these charge variations are conserved in time, that is, $\delta \dot{\mathbb{Q}}^\pm = 0$ on-shell. However, obtaining the explicit form of the charges requires integrating \eqref{eq:VSdeltaQ}, which in turn demands solving  for $\tilde{\eta}_\pm$. For generic functionals $H^\pm$, this integration problem is highly nontrivial \cite{perez2016boundary}.

In the special case where the Hamiltonians are linear in the fields, i.e. when the chemical potentials $\mu_\pm$ are constant, equation \eqref{eq:epsiloneqn} shows that the gauge transformation parameters become field-independent. As a result, the conserved charges can be integrated straightforwardly, yielding
\begin{equation}
\label{eq:VSQ}
    \mathbb{Q}^\pm[\tilde\eta_\pm] = \mp\frac{\mathrm{k}}{4\pi} \int d\theta \ \tilde\eta_\pm \, \mathcal{L}_\pm.
\end{equation}
As before, the algebra of the global charges follows directly from \eqref{eq:RGformula}. Employing the transformation law of the dynamical fields in \eqref{eq:Lgaugetransform}, this relation reduces to two independent copies of the Virasoro algebra with central extension \eqref{eq:virasoroalgebra_modes}.

One may also lift the restriction to constant chemical potentials giving rise to a broader family of boundary conditions in which the chemical potentials are allowed to depend on the boundary fields \cite{henneaux2013chemical, bunster2014generalized}. In this case, the asymptotic symmetry algebra reduces to an abelian structure, ensuring integrability.

A different simple choice of the Hamiltonians would be setting
\begin{equation}
	H^\pm = \frac{\mathrm{k}}{4 \pi} \frac{\mathcal{L}_\pm^2}{2}.
\end{equation} In this case, the field equations \eqref{eq:VSeom} for each sector turn out to be identical to the KdV equation. This is an integrable system with known solutions, given in terms of Gelfand--Dikii polynomials \cite{perez2016boundary}. In fact, we get the tower of charges \eqref{towerofcharges} with $p_\pm$ replaced by $\Ell_\pm$. Getting back the same hierarchy as in the KM case is a consequence of the bi-Hamiltonian character of the system. This holds even when the system is promoted to the Gardner hierarchy (of which KdV is a special case).

There are other generalizations of the VS boundary conditions, such as adding supersymmetry \cite{henneaux2012super, valcarcel2019new}, and higher spin fields \cite{campoleoni2010asymptotic, campoleoni2011asymptotic, henneaux2013chemical, krishnan2017chiral, henneaux2010nonlinear, ozer2020exploring, ozer2022mathcal, ozer2025n}.\footnote{The similar analysis for three dimensional asymptotically flat theories can be found in \cite{barnich2014asymptotic, barnich2017super, grumiller2017most, lodato2016super, banerjee2017n, banerjee2019maximally}. Flat space analogs of the KM boundary conditions are discussed in \cite{ammon2017higher, grumiller2020spacetime}.}

\subsection{Relating the boundary conditions}

It was shown in \cite{afshar2016soft} that the Kac--Moody and Virasoro boundary conditions are connected by a large gauge transformation. As this transformation is non-trivial at the boundary, it modifies the asymptotic form of the fields, mapping a Virasoro configuration \eqref{eq:apmthetat} into a Kac--Moody configuration \eqref{eq:apmform}. In particular, it mixes the roles of the dynamical fields and the chemical potentials, providing an explicit dictionary, referred to as the \emph{Miura map}, between the two descriptions. As a consequence, the asymptotic symmetry algebra also changes: the Virasoro algebra of the VS boundary conditions is mapped to the Kac--Moody algebra of the KM boundary conditions. Moreover, it can be shown that correspondence extends to the Hamiltonians, with the constant chemical potential (also known as Brown-Henneaux) Hamiltonian on the VS side reducing under the transformation to the relativistic free fermion Hamiltonian on the KM side. 

In this section we briefly review this correspondence. The goal is to determine a suitable gauge transformation $g$ that relates the VS gauge field $\tilde{a}^\pm$ to the KM gauge field $a^\pm$ such that
\begin{equation}
    \tilde{a}^\pm = g^{-1} \left(d + a^\pm \right) g.
\end{equation}
We perform a Gauss decomposition of the $\slr$ group element $g$ such that,
\begin{equation}
	g = \exp(x \, L_{1}) \cdot \exp(y \, L_{0}) \cdot \exp(z \, L_{-1}),
\end{equation} where the transformation parameters $x$, $y$, and $z$ are some functions of asymptotic coordinates $(t, \theta)$. 
By comparing the coefficients of each generator, one obtains the constraints on the parameters,
\begin{equation}
  \begin{split}
    x' &= e^{-y} + p_\pm \, x,\\
    y' &= -(2z + p_\pm),\\
    z' &= z^2 - \tfrac{1}{2}\mathcal{L}_\pm,
  \end{split}
\qquad
  \begin{split}
    \dot{x} &= \mu_\pm e^{-y} + \xi_\pm \, x,\\
    \dot{y} &= -(2 \mu_\pm \, z + \mu'_\pm + \xi_\pm),\\
    \dot{z} &= z (\mu_\pm \, z + \mu'_\pm) - \tfrac{1}{2}(\mu''_\pm - \mathcal{L}_\pm \,\mu_\pm).
  \end{split}
\end{equation}

In general, there are many possible choices of the functions $x$, $y$, and $z$ that satisfy the above relations and thus provide a mapping between the pairs $(\Ell_\pm, \, \mu_\pm)$ and $(p_\pm, \,\xi_\pm)$. 
A particularly simple choice, adopted in~\cite{afshar2016soft}, is
\begin{equation}
    y = 0, \qquad z = -\frac{1}{2} p_\pm,
\end{equation}
which leads to the conditions on $x(t, \theta)$:
\begin{equation}
\dot{x} - \xi_\pm \, x = \mu_\pm, 
\qquad 
x' - p_\pm \, x = 1 .
\end{equation}

The Chern--Simons equations of motion, together with the above constraints, imply that the chemical potentials are related by,
\begin{equation}
\label{chempotbosoniceqn}
\mu'_\pm - p_\pm \, \mu_\pm = -\xi_\pm.
\end{equation}

As a consequence, one finds the relation between the dynamical fields:
\begin{equation}\label{eq:Miura}
\mathcal{L}_\pm = \frac{1}{2} p^{2}_\pm + p^{\prime}_\pm .
\end{equation}
This last relation is known as the \emph{Miura transformation} \cite{miura1968korteweg}, and it provides the explicit mapping from the Virasoro algebra to the Kac--Moody algebra \cite{gardner1971korteweg}. Using the Miura map, it is easy to verify that
\begin{equation}
	\delta \mathbb{Q}^\pm = \mp \frac{\mathrm{k}}{4\pi} \oint d \theta \ \tilde{\eta}_\pm \, \delta \Ell_\pm = \mp\frac{\mathrm{k}}{4\pi} \oint d \theta \ \eta_\pm \, \delta p_\pm,
\end{equation} that is, the variation of the global charges on the VS side map \eqref{eq:VSdeltaQ} to those on the KM side \eqref{eq:KMQvar} provided the gauge transformation parameters $\tilde{\eta}_\pm$ and $\eta_\pm$ satisfy an analog of \eqref{chempotbosoniceqn} with the appropriate counterparts identified. Explicitly,
\begin{equation}
\label{gtbosoniceqn}
\tilde{\eta}'_\pm - p_\pm \, \tilde{\eta}_\pm = -\eta_\pm.
\end{equation}

To relate Hamiltonians on either side, one thus has to make the appropriate choice of gauge transformation parameters consistent with \eqref{gtbosoniceqn}. For example, we take the simple case of constant gauge transformation parameters $\tilde{\eta}_\pm$ on the VS side. Then, the charges easily functionally integrate to,
\begin{equation}
\label{eq:VSQlinearL}
	\mathbb{Q}^\pm_\text{VS} = \mp \frac{\mathrm{k}}{4\pi} \, \tilde{\eta}_\pm \oint d \theta \  \Ell_\pm.
\end{equation} 

On the KM side, however, the corresponding gauge parameter $\eta_\pm$ depends linearly on the field $p_\pm$, which allows one to integrate $\delta Q^\pm$ to obtain
\begin{equation}
	\mathbb{Q}^\pm_\text{KM} = \mp\frac{\mathrm{k}}{4\pi} \, \tilde{\eta}_\pm \oint d \theta \ \frac{p^2_\pm}{2}.
\end{equation}

This result is fully consistent with \eqref{eq:VSQlinearL}, as substituting the Miura map \eqref{eq:Miura} into the VS expression reproduces the KM charge, with the additional term being a total derivative.

In the next sections, we put together the above constructions in the presence of fermionic excitations in the boundary conditions.

\section{A minimal supersymmetric extension of Kac--Moody boundary conditions}
\label{sec:superMiura}

In this work, we investigate supersymmetric extensions of the Kac--Moody boundary conditions. In general, gravity in three spacetime dimensions with negative cosmological constant and $\N = (N,M)$ supersymmetry can be cast as a Chern--Simons theory with gauge group $\text{OSp}(N|2,\rr) \otimes \text{OSp}(M|2,\rr)$ \cite{achucarro1986chern, banados1998anti}.\footnote{Note that $\text{OSp}(0|2,\rr) \cong \slr$, thus the analysis reduces to the bosonic case correctly. For more general formulations with other gauge groups, see \cite{henneaux2000asymptotic}.}

 Here we restrict to the minimal case of $\mathcal{N}=(1,1)$ supersymmetry. The bulk gauge algebra is then $\osp(1|2,\rr)\oplus \osp(1|2,\rr)$, and each chiral sector involves boundary connections carrying both bosonic and fermionic components.
 
\subsection{Motivating the choice of gauge connection}

In this section, we are motivated to construct boundary conditions that can be related to the super--Virasoro boundary conditions introduced in \cite{banados1998anti, bautier2001ads, valcarcel2019new}
\begin{equation}
\label{eq:super-brown-henneaux}
\begin{split}
	\tilde{a}_\theta^\pm &= L_{\pm1} - \frac{1}{2} \mathcal{L}_\pm L_{\mp 1} - \frac{1}{4} \mathcal{Q}_\pm Q_{\mp 1/2}, \\
	\tilde{a}_t^\pm &= \mu_\pm \, \tilde{a}_\theta^\pm \mp \mu' L_0 + \bigg(\frac{1}{2}\mu''_\pm - \frac{1}{4} \mathcal{Q}_\pm \nu_\pm \bigg) L_{\mp1} + \nu_\pm Q_{\pm1/2} \mp \nu'_\pm Q_{\mp 1/2},
\end{split}
\end{equation} where the generators span the $\osp(1|2, \rr)$ algebra, and $\Q_\pm \param$ and $\nu_\pm \param$ denote the fermionic dynamical fields and chemical potentials respectively. The chemical potentials are typically taken to be constant at the boundary, with $\mu_\pm = -1$ and $\nu_\pm = 0$. The $\N = (1,1)$ and extended supersymmetric cases were analyzed in \cite{banados1998anti} and \cite{henneaux2000asymptotic, valcarcel2019new}, respectively.  In each of these cases, the asymptotic symmetry algebra corresponds to two copies of the relevant superconformal algebra \eqref{eq:N=1superconformal}.

Accordingly, we consider the general form of the connections $a^\pm$ as
\begin{equation}
\label{susy_bc}
\begin{split}
	a_\theta^\pm \param &= \pm \bigg(p_\pm \param \, L_0 + \frac{1}{2} \phi_\pm \param \, Q_{1/2} \bigg), \\
	a_t^\pm \param &= \xi_\pm \param \, L_0 + \frac{1}{2} \zeta_\pm \param \, Q_{1/2} + \frac{1}{2} \psi_\pm \param \, Q_{-1/2}.
\end{split}
\end{equation} Here, $p_\pm \param$ are the dynamical fields associated with the gravitational sector, while $\phi_{\pm} \param$ represent their supersymmetric counterparts. The temporal components of the connection involve the functions $\xi_\pm \param$, $\zeta_\pm \param$ and $\psi_\pm \param$, which, as we shall see below, are not all independent.\footnote{One would expect two chemical potentials corresponding to two dynamical fields. Yet, we introduce three functions in the temporal components of the connections. The role of the third component is made apparent at the end of section \ref{sec:skdv}.}

The equations of motion for the Chern-Simons action, given by \eqref{eq:Maxeq}, imply that the dynamical fields and chemical potentials satisfy,
\begin{align}
\label{eq:singlefermion_eom_p}
	\dot{p}_\pm &= \pm \xi_\pm' - \frac{1}{2}\phi_\pm \psi_\pm, \\
\label{eq:singlefermion_eom_phi}
	\dot{\phi}_\pm &= \pm \zeta_\pm' - \frac{1}{2}p_\pm \zeta_\pm + \frac{1}{2}\xi_\pm \phi_\pm,
\end{align}
along with the constraints,
\begin{align}
\label{eq:zetaconstraint}
	\phi_\pm \zeta_\pm &= 0, \\
\label{eq:psiconstraint}
	\psi'_\pm \pm \frac{1}{2}p_\pm \psi_\pm &= 0.
\end{align}
The constraints fix $\zeta_\pm \param$ and $\psi_\pm \param$ in terms of the dynamical fields. In particular, we find that $\zeta_\pm$ must be proportional to $\phi_\pm$. We denote $c^\pm[p_\pm, p'_\pm, \ldots , \phi_\pm, \phi_\pm',\ldots ]$ to be the constants of proportionality, which may be functionals of $p_\pm$, $\phi^\alpha_\pm$ or their derivatives.

The equations of motion can then be written as
\begin{align}
	\dot{p}_\pm &= \pm \xi'_\pm - \frac{1}{2}\phi_\pm \psi_\pm, \\
	\dot{\phi}_\pm &= \pm c^\pm \phi'_\pm + \frac{1}{2} (\pm 2 {c_\alpha^\pm}' - c^\pm p_\pm + \xi_\pm) \phi_\pm,
\end{align} along with the constraint \eqref{eq:psiconstraint} on $\psi_\pm$.

This is the first hint at non-triviality of this class of supersymmetric systems: the fermionic chemical potentials cannot be set arbitrarily as functions of the asymptotic coordinates and boundary fields. Note that henceforth the constraints \eqref{eq:zetaconstraint} and \eqref{eq:psiconstraint} are assumed whenever we refer to \eqref{susy_bc}.

\subsection{Canonical generators and asymptotic symmetry algebra}

As before, the variation of the gauge connection under transformations that preserve the form of \eqref{susy_bc} at the boundary is given by \eqref{eq:gtr}.
We expand the gauge parameter as\footnote{This is not the most general transformation parameter, as we have turned off generators along the bosonic directions $L_{\pm 1}$, in accordance to the bosonic case as well as the form of $a_\theta$ given in \eqref{susy_bc}. Thus, we consider only a subset of all transformations that obey the requirement.}
\begin{equation}
	\lambda^\pm = \eta_\pm L_0 + \chi_\pm^\alpha Q_\alpha,
\end{equation} where $\alpha = \pm \tfrac{1}{2}$. Thus, the variation of the dynamical fields is given by,
\begin{align}
	\delta p_\pm &= \pm \eta'_\pm - \phi_\pm \chi_\pm^{-1/2}, \\
	\delta \phi_\pm &= \pm 2\chi_\pm'^{\, 1/2} - p_\pm \chi_\pm^{1/2} + \frac{1}{2} \phi_\pm \eta_\pm.
\end{align} along with the constraints,
\begin{align}
\label{eq:chi_constraint1}
	\phi_\pm \chi_\pm^{1/2} &= 0, \\
\label{eq:chi_constraint2}
	\pm 2 \chi_\pm'^{\,-1/2} + p_\pm \chi_\pm^{-1/2} &= 0.
\end{align} 

The corresponding equations for the chemical potentials are obtained simply by replacing $p_\pm \to \xi_\pm$, $\phi_\pm \to \zeta_\pm$ , and $\theta \to t$ in the above, and adding $\psi_\pm$ as
\begin{align}
	\delta \xi_\pm &= \dot{\eta}_\pm - \zeta_\pm \chi_\pm^{-1/2} - \psi_\pm \chi_\pm^{1/2},
\end{align} along with the constraints (recall that the field-dependence of $\zeta_\pm$ and $\psi_\pm$ is already known),
\begin{align}
	\delta \zeta_\pm &= 2 \dot{\chi}^{1/2}_\pm - \xi_\pm \chi^{1/2}_\pm + \frac{1}{2} \zeta_\pm \eta_\pm, \\
	\delta \psi_\pm &= 2 \dot{\chi}^{-1/2}_\pm + \xi_\pm \chi^{-1/2}_\pm - \frac{1}{2} \psi_\pm \eta_\pm,
\end{align}
and,
\begin{equation}
	\psi_\pm \chi^{-1/2}_\pm = 0. \\
\end{equation}

Next, we compute the asymptotic symmetry algebra for the new boundary conditions. As before, the global charges \eqref{conservedcharge} satisfy,
\begin{align}
	\cancel{\delta} \mathbb{Q}^\pm [\lambda^\pm] &= - \frac{\mathrm{k}}{2 \pi} \oint d \theta \, \STr[\lambda^\pm \delta a^\pm_\theta] \\
	&= \mp \frac{\mathrm{k}}{4 \pi} \oint d \theta \, \big( \eta_\pm \delta p_\pm -2 \chi^{-1/2}_\pm\delta \phi_\pm \big)
\label{conservedcharge_susy}
\end{align}

In this case, however, the gauge parameters $\chi^{-1/2}_\pm$ that appear in the expression of the conserved charge \textit{cannot} be taken as independent of the dynamical fields, nor are they expressible in terms of independent chemical potentials, and thus, we cannot directly integrate out \eqref{conservedcharge_susy}. However, the previous section provides us a hint at an alternative approach to make $\cancel{\delta} \mathbb{Q}$ integrable.

We demand that the supersymmetric KM boundary conditions \eqref{susy_bc} we have chosen be related to the supersymmetric extension of VS boundary conditions \eqref{eq:super-brown-henneaux} by some large gauge transformation. We show the calculations for the chiral sector and drop the $\pm$ notation.
Analogous to the bosonic case, we perform a gauge transformation on the our gauge connection of the form,
\begin{equation}
    g = \exp{(c Q_\frac{1}{2})} \exp{( xL_{1})} \exp{(y L_{0})} \exp{(z L_{-1})} \exp{(d Q_{-\frac{1}{2}})},
\end{equation}where $c,d$ are the Grassmann-valued gauge transformation parameters.

The most general map is given by the constraints
\begin{align}
        x' &= e^{-y} + xp + c(e^{-\frac{1}{2}y} d - \frac{1}{2}\phi), \\
        y' &= -2z -p, \\
        z' &= z^2 - \frac{1}{2}\mathcal{L} + \frac{1}{4} d\mathcal{Q}, \\
        c' &= \frac{1}{2}cp - e^{-\frac{1}{2}y} d  - \frac{1}{2}\phi, \\
        d' &= dz - \frac{1}{4} \mathcal{Q},
\end{align} and,
\begin{align}
        \dot{x} &= \mu e^{-y} + x (\xi + \frac{1}{2}c\psi) + c(e^{-\frac{1}{2}y} (d \mu - \nu) - \frac{1}{2}\zeta), \\
        \dot{y} &= -2(z \mu + \frac{\mu'}{2} + d\nu + \frac{\xi}{2} - + \frac{1}{2}c\psi), \\
        \dot{z} &= z^2 \mu + z\mu' + (\frac{1}{2}\mu'' - \frac{1}{2}\mathcal{L}\mu - \frac{1}{4}\mathcal{Q}\nu) + d \big(\nu'  + \frac{1}{4} \mu \mathcal{Q}+z\nu - \frac{1}{2} e^{-\frac{1}{2}y} \psi \big) -e^{-y}, \\
        \dot{c} &= \frac{1}{2} c \xi - e^{-\frac{1}{2}y} (d\mu - \nu) - \frac{1}{2}\zeta + \frac{1}{2} x \psi, \\
        \dot{d} &= d\big(z\mu + \frac{\mu'}{2}\big) - \big(\nu' + \frac{1}{4} \mu \mathcal{Q} + z\nu \big) - \frac{1}{2} e^{-\frac{1}{2}y} \psi.
\end{align} 
As before, we take a special case to simplify the analysis. Taking
\begin{equation}
	y=0, \qquad z = -\dfrac{1}{2}p, \qquad c = 0, \qquad d = -\dfrac{1}{2}\phi,
\end{equation} we find that $x \param$ must fulfil, \begin{align}
        x' &= 1 + xp, \\
        \dot{x} &= \mu + x \xi,
\end{align} as before. This gives us the relation between the dynamical fields to be
\begin{align}
\label{N=1_fields_map}
        \mathcal{L} &= \frac{1}{2}p^2 +p' - \frac{1}{2}\phi \phi', \\
        \mathcal{Q} &= p \phi + 2 \phi',
\end{align} which is known as the \textit{super-Miura transformation} \cite{mathieu1988superconformal, mathieu1988supersymmetric}, and those between the chemical potentials to be,
\begin{align}
\label{chempotrel1}
	\mu' - p \mu + \nu \phi &= - \xi, \\
\label{chempotrel2}
	\nu' - \frac{1}{2}p\nu + \mu \big(\frac{1}{4}p \phi + \frac{1}{2}\phi'\big) &= \frac{1}{2} \big(\zeta' -\frac{1}{2}p \zeta + \xi \phi \big) - \frac{1}{2}\psi.
\end{align}

It is then straightforward to verify that,
\begin{equation}
\label{eq:delta_Q_susy}
	\delta \mathbb{Q}^\pm = \mp \frac{\mathrm{k}}{4\pi} \oint d \theta \big( \tilde{\eta}_\pm \, \delta \Ell_\pm - \tilde{\chi}_\pm \, \delta \Q_\pm \big) = \mp \frac{\mathrm{k}}{4\pi} \oint d \theta \big(\eta_\pm \,\delta p_\pm - 2 \chi^{-1/2}_\pm \, \delta \phi_\pm \big),
\end{equation} where $\tilde{\chi}_\pm$ is the gauge transformation parameter along $Q_{1/2}$ in the super-Virasoro analysis, provided the gauge transformation parameters satisfy \eqref{chempotrel1} and \eqref{chempotrel2} under the correct identifications. Finally, we use these relations to redefine the gauge transformation parameters on the super-Kac-Moody side to get the RHS as,
\begin{equation}
    \delta \mathbb{Q}^\pm = \mp \frac{\mathrm{k}}{4 \pi} \oint d \theta \bigg[ \big(\tilde{\eta}_\pm p_\pm - \tilde{\eta}'_\pm - \tilde{\chi}_\pm \phi_\pm \big) \delta p_\pm + \bigg(\frac{1}{2} \tilde{\eta}_\pm \phi'_\pm + \frac{1}{2} (\tilde{\eta}_\pm \phi_\pm)' - \tilde{\chi}_\pm p_\pm + 2 \tilde{\chi}'_\pm \bigg) \delta \phi_\pm \bigg].
\end{equation}

For parameters $\tilde{\eta}_\pm$ and $\tilde{\chi}_\pm$ independent of the dynamical fields $p_\pm$ and $\phi_\pm$, this integrates out to,
\begin{equation}
\label{eq:integrated_Q}
	\mathbb{Q}^\pm = \mp \frac{\mathrm{k}}{4 \pi} \oint d\theta \bigg( \frac{1}{2} \tilde{\eta}_\pm p^2_\pm - \tilde{\eta}'_\pm p_\pm - \frac{1}{2} \tilde{\eta}_\pm \phi_\pm \phi'_\pm - \tilde{\chi}_\pm p_\pm \phi_\pm + 2 \tilde{\chi}'_\pm \phi_\pm \bigg).
\end{equation}

Using the Regge-Teitelboim formula \eqref{eq:RGformula}, we get the Poisson structure,
\begin{align}
\label{eq:pp_Poisson}
\{p_\pm (t,\theta), p_\pm (t,\theta') \} &= \mp \frac{4 \pi}{\mathrm{k}} \partial_\theta \delta(\theta - \theta'), \\
\label{eq:phiphi_Poisson}
\{\phi_\pm (t,\theta), \phi_\pm (t,\theta') \} &= \pm \frac{4 \pi}{\mathrm{k}} \delta(\theta - \theta').
\end{align}

The above relations clearly show that the bosonic and fermionic symmetry generators completely decouple for this choice of the boundary conditions.

\subsection{Boundary Hamiltonians and the super-KdV hierarchy}
\label{sec:skdv}

As in the bosonic case, one has freedom to choose the form of the independent chemical potentials (or equivalently the boundary Hamiltonian) and this choice dictates the boundary dynamics of the system. In this section, we illustrate this with a particularly simple choice. Let us set $\zeta_\pm \equiv -\phi_\pm$ (i.e. $c ^\pm = -1$) and $\xi_\pm = -p_\pm$ for all $\theta$, $t$ at the asymptotic boundary. We also set $\psi_\pm = 0$ consistent with \eqref{eq:psiconstraint}.

Under this condition, we find that the dynamical fields and the chemical potentials satisfy the simplified equations of motion \eqref{eq:singlefermion_eom_p} and \eqref{eq:singlefermion_eom_phi}:
\begin{align}
\label{eq:simple_eom_1}
	\dot{p}_\pm  &= \mp p_\pm', \\ \label{eq:simple_eom_2}
	\dot{\phi}_\pm &= \mp \phi'_\pm.
\end{align}

This also further constraints the gauge transformation parameters as,
\begin{equation}
\label{eq:simple_constraints}
	\dot{\eta}_\pm = \mp \eta'_\pm, \qquad \dot{\chi}_\pm^{\, \alpha} = \mp {\chi^\alpha_\pm}'.
\end{equation}

To construct the boundary Hamiltonian corresponding to \eqref{eq:simple_eom_1} and \eqref{eq:simple_eom_2} is simple. Using the Poisson structure \eqref{eq:pp_Poisson} and \eqref{eq:phiphi_Poisson}, one finds a suitable choice for the Hamiltonians $H^\pm$ to be
\begin{equation}
	H^\pm = \frac{\mathrm{k}}{4 \pi} \int d \theta \frac{1}{2} \Big(p^2_\pm - \phi_\pm \phi'_\pm \Big).
\end{equation}

Note that the supersymmetric generalization of \eqref{eq:deltaBinfty} does not include a fermionic term for this special case as we have set $\psi_\pm$ to zero. Thus, the functionals $H^\pm$ cannot be motivated as the boundary terms in each sector of the action, unlike the bosonic case. 

However, they can be shown to be conserved charges of the gauge symmetry that preserves the form of the boundary conditions. Namely, choosing $\tilde{\eta}_\pm = -1$ and $\tilde{\chi}_\pm = 0$ (which is equivalent to choosing $\mu_\pm = -1$ and $\nu_\pm = 0$ in the original formulation of super-VS boundary conditions) in \eqref{eq:integrated_Q}, we observe that the above $H^\pm$ satisfy 
\begin{equation}
	\mathbb{Q}^\pm = \pm H^\pm.
\end{equation}

Equivalently, this could be seen as choosing the gauge transformation parameters,
\begin{equation}
	\eta_\pm = -p_\pm, \qquad \chi^{1/2}_\pm = -\frac{1}{2} \phi_\pm, \qquad \chi^{-1/2}_\pm = 0,
\end{equation} which satisfy the constraints \eqref{eq:simple_constraints} and the previous constraints \eqref{eq:chi_constraint1} and \eqref{eq:chi_constraint2}. 

Note that this looks strange. Inserting these gauge transformation parameters in \eqref{eq:delta_Q_susy} would yield a purely bosonic conserved charge. However, here we have a supersymmetric one. This is a consequence of the sleight of hand in the previous section. By keeping $\psi_\pm$ instead of setting it to zero at the beginning, we are able to use \eqref{chempotrel2} to integrate $\delta \mathbb{Q}^\pm$ while preserving the fermionic terms.

Finally, we comment on other conserved charges of the system. Similar to the bosonic case \eqref{towerofcharges}, it is possible to construct a tower of charges, corresponding to different choices of the parameters, now taken to be field-dependent, with each relating to an element of the \emph{super--KdV} or \emph{super--Gardner} hierarchy \cite{mathieu1988supersymmetric}. The first few charges are given by,
\begin{equation}
\begin{split}
		H^\pm_0 &= \frac{\mathrm{k}}{4\pi}\int d\theta \, p_\pm, \\
		H^\pm_1 &= \frac{\mathrm{k}}{4\pi}\int d\theta \, \frac{1}{2} \big(p_\pm^2 - \phi_\pm \phi'_\pm \big), \\
		H^\pm_2 &= \frac{\mathrm{k}}{4\pi}\int d\theta \, \frac{1}{3} \big(p_\pm^3 + p_\pm'^{\,2} - 2 p_\pm \phi_\pm \phi'_\pm - \frac{1}{2} \phi'_\pm \phi''_\pm \big), \\
		H^\pm_3 &= \frac{\mathrm{k}}{4\pi}\int d\theta \, \frac{1}{4} \big(p_\pm^4 + 4p_\pm p_\pm'^{\,2} + \frac{4}{5} p_\pm''^{\, 2} - 3 p^2_\pm \phi \phi' + \frac{2}{5} p_\pm \phi'_\pm \phi'' + \frac{8}{5} p_\pm \phi_\pm \phi'''_\pm - \frac{1}{5} \phi''_\pm \phi'''_\pm \big). \\
	\end{split}
\end{equation}

It is straightforward, if tedious, to verify that these charges are all in involution.

\section{A different fermionic extension of Kac--Moody boundary conditions}
\label{sec:superKM}

In the previous section, we developed a supersymmetric extension of the Kac--Moody boundary conditions that can be mapped to the super--Virasoro boundary conditions through a large gauge transformation. However, the resulting Poisson structure is trivial, as it factorizes into independent bosonic and fermionic sectors. The corresponding boundary field theory remains supersymmetric. In this section, we introduce a more intricate fermionic extension of the Kac--Moody boundary conditions, still focussing on the minimal $\mathcal{N}=(1,1)$ case. 

\subsection{Poisson structure and large gauge symmetry}

This time, we consider the more general boundary connection
\begin{equation}
\label{eq:generic_bc_n=1}
	a^\pm = (\xi_\pm \, d t \,\pm\, p_\pm \, d \theta) L_0^\pm \,+\, \frac{1}{2}(\zeta_\pm^\alpha \, d t \,\pm\, \phi_\pm^\alpha \, d \theta)Q^{\pm}_\alpha,
\end{equation} where $\alpha = \pm \tfrac{1}{2}$, summed over as before.

Here, we extend the bosonic Kac-Moody boundary ansatz by activating both fermionic components $\phi^\alpha_\pm$ and their corresponding chemical potentials $\zeta^\alpha_\pm$ along the fermionic generators, thereby incorporating the full fermionic sector into the boundary data.

The Chern-Simons equations of motion turn out to be
\begin{align}
	\dot{p}_\pm &= \pm \xi'_\pm - \tfrac{1}{2}\phi^\alpha_\pm \zeta^{-\alpha}_\pm, \\
	\dot{\phi}^\alpha_\pm &= \pm \zeta'^\alpha_\pm + \alpha \phi^\alpha_\pm \xi_\pm - \alpha p_\pm \zeta^\alpha_\pm,
\end{align} along with the constraints,
\begin{equation}
\label{eq:zeta_constraint_main}
	\phi^\alpha_\pm \zeta^\alpha_\pm = 0,
\end{equation} where $\alpha$ is \emph{not} summed over. These constraints are satisfied by taking the fermionic chemical potentials $\zeta^\alpha_\pm$ to be proportional to the fermionic fields $\phi^\alpha_\pm$. Like before, we denote the constants of proportionality with $c_\alpha^\pm$, which may be functionals of $p_\pm$, $\phi^\alpha_\pm$ or their derivatives, or other functions of spacetime. Henceforth, \eqref{eq:zeta_constraint_main} is assumed whenever we refer to \eqref{eq:generic_bc_n=1}.

The equations of motion reduce to
\begin{align}
	\dot{p}_\pm &= \pm \xi'_\pm - \tfrac{1}{2} c^\pm_\alpha \phi^\alpha_\pm \phi^{-\alpha}_\pm, \\
	\dot{\phi}^\alpha_\pm &= \pm c_\alpha^\pm \phi'^{\, \alpha}_\pm + \alpha (\pm 4 \alpha {c_\alpha^\pm}' - c_\alpha^\pm p_\pm + \xi_\pm ) \, \phi^\alpha_\pm.
\end{align} 

For the generic boundary condition \eqref{eq:generic_bc_n=1}, the boundary term \eqref{eq:deltaBinfty} is given by
\begin{equation}
	\delta \mathcal{B}_\infty^\pm \,=\, \mp \,\frac{\mathrm{k}}{4 \pi} \int_{\partial \mathcal{M}} dt \wedge d\theta \,\; \big(\xi_\pm \, \delta p_\pm \,+\, 2\alpha \,\zeta^\alpha_\pm \, \delta \phi^{-\alpha}_\pm \big),
\end{equation} where $\alpha$ is summed over

As in the bosonic case, the existence of a boundary term in the action suggests introducing functionals $H^\pm$, which may depend on the dynamical fields $p_\pm$, $\phi^\alpha_\pm$ and their derivatives, and other spacetime-dependent functions.
\begin{equation}
	H^\pm = \int d \theta \; \mathcal{H}^\pm [p_\pm, p'_\pm, p''_\pm, \ldots, \phi^\alpha_\pm, {\phi^\alpha_\pm}', {\phi^\alpha_\pm}'', \ldots].
\end{equation}

We could choose the chemical potentials to be
\begin{equation}
\label{eq:Chpot}
	\xi_\pm = - \frac{4 \pi}{\mathrm{k}} \fdel{H^\pm}{p_\pm}, \qquad
	\zeta_\pm^\alpha = -2\alpha \frac{4 \pi}{\mathrm{k}} \fdel{H^\pm}{\phi^{-\alpha}_\pm} = c_\alpha^\pm \phi^\alpha_\pm,
\end{equation} where $\alpha$ is \emph{not} summed over in the last equality. Then, the corresponding boundary term is given by \eqref{eq:Binfty}, and the dynamics of the fields depends on the choice of the boundary conditions as
\begin{align}
	\dot{p}_\pm &= -\frac{4 \pi}{\mathrm{k}} \bigg(\pm \partial_\theta \bigg( \fdel{H^\pm}{p_\pm} \bigg) + \alpha \phi^\alpha_\pm \bigg(\fdel{H^\pm}{\phi^{\alpha}_\pm} \bigg) \bigg), \\ 
	\dot{\phi^\alpha_\pm} &= - \frac{4 \pi}{\mathrm{k}} \bigg( \alpha \, \phi^\alpha_\pm \bigg(\fdel{H^\pm}{p_\pm}\bigg) +\big(\pm 2 \alpha \,\partial_\theta - \tfrac{1}{2} \, p_\pm \big) \bigg(\fdel{H^\pm}{\phi^{-\alpha}_\pm} \bigg) \bigg).
\end{align}

Using this, one can derive the non-trivial Poisson brackets of the system for arbitrary $H^\pm$ to be
\begin{equation}
\label{eq:PoissonB}
\begin{split}
	\{p_\pm (t,\theta), p_\pm (t,\theta') \}_\text{PB} &= \mp \, \frac{4 \pi}{\mathrm{k}} \, \delta'(\theta - \theta'), \\
	\{p_\pm (t,\theta), \phi^\alpha_\pm (t,\theta') \}_\text{PB} &= \pm \, \alpha \frac{4 \pi}{\mathrm{k}} \, \phi_\pm^\alpha (t,\theta) \, \delta(\theta - \theta'), \\
	\{\phi^\alpha_\pm (t,\theta), \phi^\beta_\pm (t,\theta') \}_\text{PB} &= \pm \,\frac{2\pi}{\mathrm{k}} \, p_\pm(t,\theta) \, \delta^{\alpha + \beta, 0} \, \delta(\theta - \theta') \pm \frac{4\pi}{\mathrm{k}} C^{\alpha \beta} \, \delta'(\theta - \theta').
\end{split}
\end{equation}

Next, we study the large gauge symmetries of the system. The boundary variation of the gauge field is given by \eqref{eq:gtr}, where we now expand the gauge transformation parameter as 
\begin{equation}
\lambda^\pm \,=\, \eta_\pm \, L_0 + \eta^1_\pm  \,L_1 + \eta^{-1}_\pm \,L_{-1} + \chi^\alpha_\pm \,Q_\alpha. 
\end{equation}

The dynamical fields vary as
\begin{align}
\label{LGT}
 \delta p_\pm \,&=\, \pm \,\eta'_\pm - \phi^\alpha_\pm \chi^{-\alpha}_\pm, \\
\label{LGT2}
    \delta \phi^{\alpha}_\pm \,&=\, \pm \, 2{\chi^\alpha_\pm}^\prime - 2\alpha p_\pm \chi^\alpha_\pm + \alpha \phi^\alpha_\pm \eta_\pm - 2 \alpha \phi^{-\alpha}_\pm \eta^{2 \alpha}_\pm.
\end{align}

Similarly the variations of the chemical potentials read as,
\begin{align}
\label{xivariation}
	\delta \xi_\pm \,&=\, \dot{\eta}_\pm - \zeta^\alpha_\pm \chi^{-\alpha}_\pm, \\
\label{zetavariation}
    \delta \zeta^\alpha_\pm \,&=\, 2\dot{\chi}^\alpha_\pm - 2\alpha \xi_\pm \chi^\alpha_\pm + \alpha \zeta^\alpha_\pm \eta_\pm - 2\alpha \zeta^{-\alpha}_\pm \eta^{2 \alpha}_\pm.
\end{align}

The parameters (for $a = \pm 1$) also satisfy the constraints,
\begin{align}
\label{eta_constraint_1}
	\eta'^{\,a}_\pm \,&=\, \pm \big(a p_\pm \eta^a_\pm  + 2 \phi^{a/2}_\pm \chi^{a/2}_\pm \big), \\
\label{eta_constraint_2}
	\dot{\eta}^a_\pm &= a \xi_\pm \eta^a_\pm  + 2 \zeta^{a/2}_\pm \chi^{a/2}_\pm.
\end{align}

To maintain consistency with the bosonic case, we set the gauge parameters along the $L_{\pm 1}$ directions to zero. This choice is compatible with the evolution equations \eqref{eta_constraint_1} and \eqref{eta_constraint_2}, provided we take $\chi^\alpha_\pm \sim \phi^\alpha_\pm$, as both are Grassmann-valued. Consequently, the fermionic gauge parameters become field-dependent, analogous to the fermionic chemical potentials. At this stage, the proportionality constants remain undetermined.

The conserved charges \eqref{conservedcharge} associated with large gauge transformations satisfy
\begin{equation}
	\delta \mathbb{Q}^\pm [\lambda^\pm] \,=\, \mp \frac{\mathrm{k}}{4 \pi} \oint d \theta \, \big( \eta_\pm \, \delta p_\pm \,+\, 4 \alpha \chi^\alpha_\pm \, \delta \phi^{-\alpha}_\pm \big).
\end{equation}

Taking $\chi^\alpha_\pm = f_\alpha^\pm \phi^\alpha_\pm$ (no sum over $\alpha$), with $f_\alpha^\pm$ a field-independent function of spacetime and equal for both $\alpha=\pm \tfrac{1}{2}$, that is, requiring that $f_\alpha^\pm \equiv f^\pm$, the variation functionally integrates to
\begin{align}
	\mathbb{Q}^\pm [\lambda^\pm] &= \mp \frac{\mathrm{k}}{4 \pi} \oint d\theta \, \big( \eta_\pm \,p_\pm \,+\, 2 f^\pm \, \phi^{1/2}_\pm \phi^{-1/2}_\pm \big).
\end{align}

Finally, one observes that the Poisson structure \eqref{eq:PoissonB} is consistent with the Regge--Teitelboim condition \eqref{eq:RGformula} and closes with an abelian structure.

\subsection{Constraints on the proportionality factors}

So far, our analysis has not imposed any constraints on the proportionality factors $c_{\alpha}^\pm$ or $f^\pm$ introduced above. However, requiring invariance of the Hamiltonian under large gauge transformations leads to specific conditions on these parameters.

To examine this, let us consider the variation of the Hamiltonian under supersymmetric large gauge transformations. We apply \eqref{eq:gtr} with only $\chi^{\alpha}$ turned on, together with \eqref{LGT} and \eqref{LGT2}. We then obtain,
\begin{align}
	\delta_\text{susy} \, p_\pm \,&=0, \\
    \delta_\text{susy} \, \phi^{\alpha}_\pm \,&=\, \pm \, 2 f^\pm {\phi^\alpha_\pm}^\prime - \big(2\alpha p_\pm f^\pm \, \mp \, 2 {f^\pm}' \big)  \phi^\alpha_\pm.
\end{align}

Using these relations, the variation of the Hamiltonian becomes
\begin{align}
\delta_\text{susy} H^\pm &= \int d\theta \,\bigg(\frac{\partial{\mathcal{H}^\pm}}{\partial p_\pm} \delta_\text{susy} \, p_\pm + \frac{\partial{\mathcal{H}^\pm}}{\partial \phi^{\alpha}_\pm} \delta_\text{susy} \, \phi^{\alpha}_\pm  \bigg) \\ 
	&= \pm \frac{\mathrm{k}}{4\pi}\int d\theta \, ( c_{-\alpha}^\pm \phi^{-\alpha}_\pm) \, \big( 4\alpha f^\pm {\phi^\alpha_\pm}^\prime \,\mp\, ( p_\pm f^\pm - 4 \alpha {f^\pm}') \, \phi^\alpha_\pm \big),
\end{align}
where the index $\alpha$ is summed over. The above expression vanishes provided the following condition holds:
\begin{equation}
\begin{split}
	 \int d\theta \bigg[{f^\pm}' \big(c^\pm_{1/2} &+ c^\pm_{-1/2} \big) \, \phi^{1/2}_\pm \phi^{-1/2}_\pm \,\pm\, \frac{1}{2} p_\pm f^\pm \big( c^\pm_{1/2} - c^\pm_{-1/2} \big) \,\phi^{1/2}_\pm \phi^{-1/2}_\pm  \\ 
	 &+ \, f^\pm \big( c^\pm_{1/2} \, \phi^{1/2}_\pm {\phi^{-1/2 \,}_\pm}' \, + \, c^\pm_{-1/2} \, {\phi^{1/2 \,}_\pm}' \phi^{-1/2 \,}_\pm \big) \bigg]= 0
\end{split}	
\end{equation}

It is straightforward to verify that this condition is satisfied for
\begin{equation}
\label{proportionalityfactors}
	c^\pm_\alpha \equiv \frac{c^\pm}{2}, \qquad {f^\pm}' = {c^\pm_\alpha}' = 0
\end{equation} for both $\alpha = \pm \tfrac{1}{2}$.

Hence, for the $\mathcal{N}=(1,1)$ supersymmetric-invariant Hamiltonian, all proportionality factors must be field-independent and independent of the angular coordinate $\theta$. These conditions also ensure that the Hamiltonian remains invariant under the full set of large gauge transformations preserving the asymptotic form of the gauge connections. 

Analogous constraints apply to the bosonic chemical potentials $\xi_\pm$ if both they and their associated gauge transformation parameters $\eta_\pm$ are assumed to be field-independent. However, in the most general case, when these quantities are allowed to be functionals of the fields, they remain unconstrained.

Finally, to determine the time dependence of the proportionality factors, we examine \eqref{xivariation} and \eqref{zetavariation}. After consistently setting $\eta^{\pm 1}_\pm = 0$, \eqref{zetavariation} further constrains $f^\pm$ to be time-independent. Thus $f^\pm$ are arbitrary constant parameters, whereas $c^\pm$ in \eqref{proportionalityfactors} may depend arbitrarily on time.

\subsection{Generic Hamiltonians, conserved charges and integrability}

We are now in a position to express the most general form of the Hamiltonian for each sector of the theory. It is given by
\begin{equation}
\label{Hamiltonian}
  H^\pm = \frac{\mathrm{k}}{4 \pi} \int_0^{2 \pi} d\theta \, \big(g^\pm(p_\pm)  +  \, c^\pm(t) \, \phi^{1/2}_\pm \phi_\pm^{-1/2} \big),
\end{equation}
where $g^\pm(p_\pm)$ denote arbitrary functions of $p_\pm$ and their derivatives, constrained only by the condition \eqref{eq:etaeq}, while $c^\pm(t)$ are arbitrary functions of time. Using the Poisson brackets \eqref{eq:PoissonB}, one can show that the following relations hold:
\begin{align}
	\{p_\pm(t,\theta), \, (\phi^{1/2}_\pm \phi^{-1/2}_\pm)(t,\theta') \}_\text{PB} \,&=\, 0, \\ 
	\int d\theta \, d\theta' \, \{ (\phi^{1/2}_\pm \phi^{-1/2}_\pm)(t,\theta), \, (\phi^{1/2}_\pm \phi^{-1/2}_\pm)(t,\theta') \}_\text{PB} \,&=\, 0.
\end{align}

Motivated by the analysis in section \ref{asa_analysis}, we now define a family of functionals:\footnote{As before, the label $n$ is arbitrary until a specifc choice of functionals is made.} 
\begin{equation}
 \mathbb{Q}_n^\pm \,=\, \pm \frac{\mathrm{k}}{4 \pi} \int_0^{2 \pi} d\theta \, \big( g_n^\pm (p_\pm)  +  c_n^\pm(t) \, \phi^{1/2}_\pm \phi_\pm^{-1/2} \big),
\end{equation}
where $g^\pm_n(p)$ are chosen in accordance with the above constraints. An explicit example of such a series is given in \eqref{towerofcharges}. It is straightforward to verify that
\begin{equation}
	\{\mathbb{Q}_m, \, \mathbb{Q}_n \}=0,
\end{equation} for any $m$ and $n$. Therefore, the functionals defined above constitute a countably infinite set of conserved charges, implying that the system is integrable.

\subsection{A free Fermi description}
\label{sec:droplet}

\begin{figure}[H]
\center
  \includegraphics[width=0.60\textwidth]{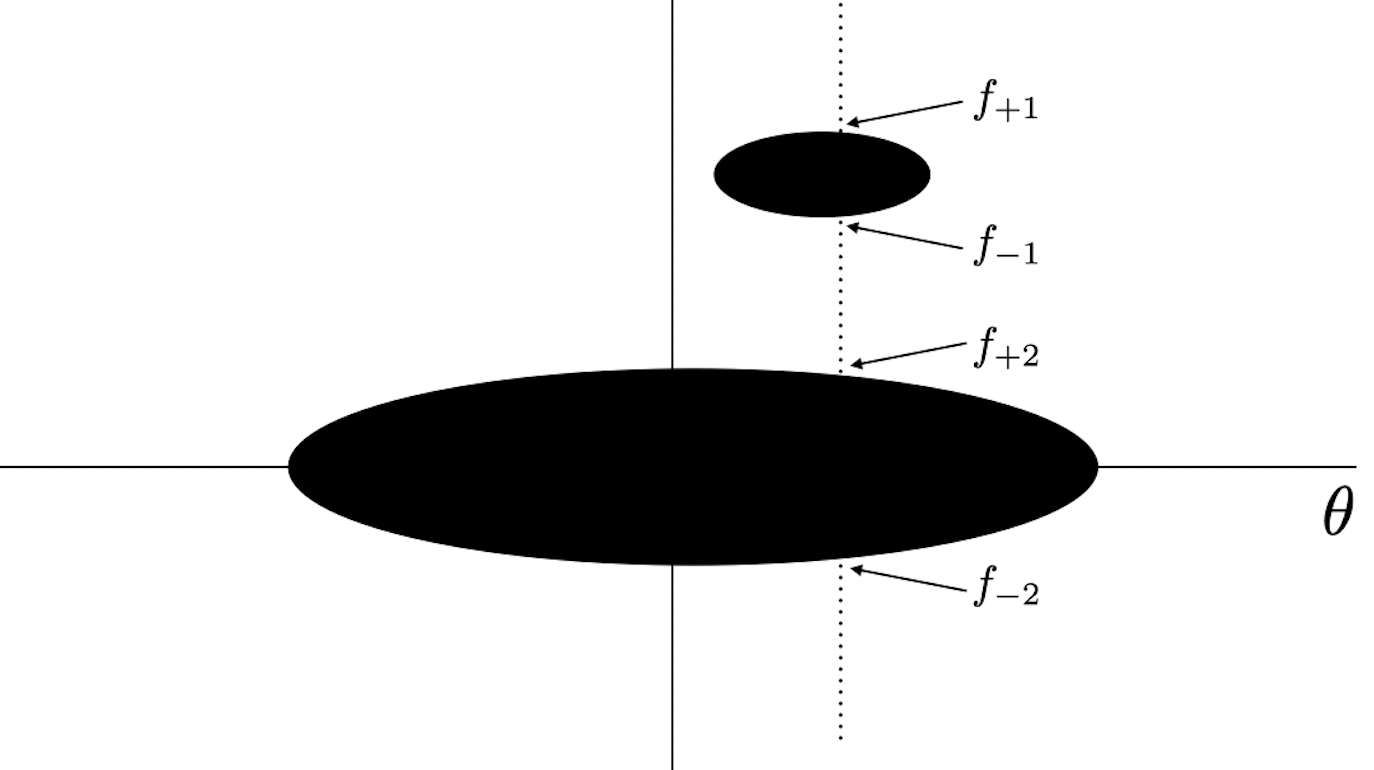}
  \caption{Folds on Fermi surface \cite{dutta2023higher}.}
  \label{fig:onefold}
  \centering
\end{figure}

As discussed in section \ref{sec:rel_free_fermi}, the collective field theory Hamiltonian admits a geometric representation in terms of fermionic droplets in phase space \cite{polchinski1991classical}. Up to this point, we have assumed that $p_\pm(t, \theta)$ are single-valued, but more generally the Fermi surface may develop \emph{folds}, where $p_\pm$ become multi-valued in $\theta$. Such folded droplets can remain attached to the main droplet or appear as disconnected components (see figure \ref{fig:onefold}). Their structure was studied in \cite{das1996folds, das2004d}, where additional fields $w_{\pm n}$ were introduced. These encode higher moments of the distribution and close a classical $w_\infty$ algebra \cite{avan1993interacting}:
\begin{equation}
\{ w_{\pm m}(t,\theta),\, w_{\pm n}(t,\theta') \} \,=\, \mp \,\frac{4\pi}{\mathrm{k}}\Big( m\,w_{\pm (m+n-1)}(t,\theta) 
+ n\,w_{\pm (m+n-1)}(t,\theta') \Big)\, 
\partial_\theta \delta(\theta-\theta').
\end{equation}

Within this framework, the phase-space integrals take the form  
\begin{align}
\int dp \ \varpi &= p_+ - p_- , \\
\int dp \ \frac{p^n}{n}\varpi &= \frac{p_+^{n+1}-p_-^{n+1}}{n(n+1)} 
+ \sum_{k=1}^n c_k^n \big(p_+^{\,n-k}w_{+k}-p_-^{\,n-k}w_{-k}\big),
\end{align}
where the coefficients $c_k^n$ are constants. For $n=1$, $2$ we obtain,
\begin{equation}
\begin{split}
    \int d\theta \, dp \ p\varpi & = \int d\theta \left[\left(\frac{p_+^2}{2} -\frac{p_-^2}{2}\right) + c_1^1 (w_{+1} -w_{-1}) \right],\\
    \int d\theta \, dp \ \frac{p^2}{2}\varpi & = \int d\theta \left[\left(\frac{p_+^3}{6} -\frac{p_-^3}{6}\right) + c_2^1 (p_+ w_{+1} - p_- w_{-1}) + c^2_2 (w_{+2} - w_{-2})\right].
\end{split}
\end{equation}

For a single fold on the upper surface (see figure \ref{fig:onefold}), the decomposition involves four turning points: $\{f_{+1},f_{-1},f_{+2},f_{-2}\}$. One may write
\begin{equation}
p_+ = f_{+1}-f_{-1}+f_{+2}, \qquad p_- = f_{-2},
\end{equation}
and the corresponding $w_{+n}$ are determined by polynomial relations among these turning points. When the fold disappears, $w_{\pm n}\to 0$ and one recovers the simple droplet with only $p_\pm$. More generally, for $T$ folds the parameterisation is  
\begin{equation}
p_+ = \sum_{i=1}^T f_{+i} - \sum_{i=1}^{T-1} f_{-i}, \qquad p_- = f_{-T},
\end{equation}
together with the moment relations
\begin{equation}
\sum_{i=1}^T \frac{f_{+i}^{n+1}-f_{-i}^{n+1}}{n(n+1)}
= \frac{p_+^{n+1}-p_-^{n+1}}{n(n+1)} 
+ \sum_{k=1}^n c_k^n \big(p_+^{\,n-k}w_{+k}-p_-^{\,n-k}w_{-k}\big).
\end{equation}  
Folded droplets therefore represent physical excitations rather than mere geometric features. Their classical dynamics is governed jointly by $p_\pm$ and $w_{\pm n}$. This picture was connected to higher-spin dynamics in $AdS_3$ \cite{dutta2023higher}. Our aim here is to show that fermionic excitations can also be captured in terms of folded Fermi surface dynamics.

In a fermionic extension, bosonic excitations naturally split into two classes: (i) purely bosonic modes, and (ii) composite states built from fermion bilinears. Both types admit a geometric description in terms of folded droplets governed by an appropriate Hamiltonian. Although the Poisson structure between $p_\pm$ and $\phi^\alpha_\pm$ is not decoupled, the equations of motion derived from the Hamiltonians
\begin{equation}
\label{eq:bndry_H}
    H^\pm = \, \frac{\mathrm{k}}{4 \pi} \oint d\theta \, \bigg( \frac{p_\pm^2}{2} -  \, \phi^{1/2}_\pm \phi_\pm^{-1/2} \bigg)
\end{equation}
are
\begin{equation}
\begin{split}
    \dot p_\pm(t,\theta) &= \mp \, p'_\pm(t,\theta),\\
    \dot \phi^\alpha_\pm(t,\theta) &= \mp \, {\phi^{\alpha \,}_\pm}'(t,\theta).
\end{split}
\end{equation}

Introducing the bilinear bosonic variable
\begin{equation}
    \mathrm{w}_\pm(t,\theta) \equiv \phi^{1/2}_\pm(t,\theta)\,\phi^{-1/2}_\pm(t,\theta),
\end{equation}
one finds that $\{\mathrm{w}_\pm, p_\pm\}_\text{PB} = 0$, while the Poisson bracket of $\mathrm{w}_\pm$ with itself is
\begin{equation}
    \{ \mathrm{w}_\pm(t,\theta), \mathrm{w}_\pm(t,\theta') \}_\text{PB} \,=\, \pm \, \frac{4\pi}{\mathrm{k}} \left(\mathrm{w}_\pm(t,\theta) + \mathrm{w}_\pm(t,\theta') \right)\delta'(\theta - \theta').
\end{equation}
In terms of $\mathrm{w}_\pm$, the boundary Hamiltonian \eqref{eq:bndry_H} becomes
\begin{equation}
    H^\pm
      = \frac{\mathrm{k}}{4\pi} \int d\theta \left( \frac{p_\pm^2}{2} - \mathrm{w}_\pm(t,\theta) \right).
\end{equation}

The equations of motion for $p_\pm(t,\theta)$ remain unchanged, while for $\mathrm{w}_\pm(t,\theta)$ one finds
\begin{equation}
    \dot{\mathrm{w}}_\pm = \mp \, \mathrm{w}'_\pm.
\end{equation}
This equation is automatically satisfied once the equations for $\phi^\alpha_\pm$ hold.

The resulting dynamics closely parallels the fold equations of relativistic fermions. The bilinears $\mathrm{w}_\pm$ (composite bosonic modes) can be identified with $w_{\pm 1}$, i.e. the first fold modes of the droplet. Importantly, if the initial state is purely bosonic, no composite excitations are generated dynamically. The fold description works for relativistic fermions because $p_\pm$ and $w_{\pm 1}$ are decoupled, and the Poisson algebra of $w_{\pm 1}$ closes on itself. Thus, starting with nonzero $p_\pm$ and $w_{\pm 1}$, higher $w_{\pm m}$ modes (i.e. $m\geq 2$) are never generated dynamically.

By contrast, if one considers the bosonic Hamiltonian as the phase-space Hamiltonian of non-relativistic fermions, the fold description breaks down. In this case $p_\pm$ and $w_{\pm 1}$ couple directly, and even if one starts with nonzero $p_\pm$ and $w_{\pm 2}$ (with all other $w_{\pm m}$ set to zero), the Poisson algebra inevitably generates the full tower of $w_{\pm m}$ modes over time.

\subsection{Quantization and soft hair}
\label{sec:quantisation}

In the microscopic understanding of black hole entropy, the \emph{hair modes} residing on the horizon play a crucial role \cite{hawking2016soft, afshar2017near, afshar2017horizon, grumiller2020near}. In three-dimensional gravity, the topological nature of the theory ensures that the same symmetry structure emerges at any constant-radial hypersurface. For asymptotically AdS$_3$ gravity, this implies that the BTZ black holes exhibit the same infinite-dimensional symmetry algebra both at the asymptotic boundary and at the event horizon. Consequently, this symmetry gives rise to an infinite set of \emph{soft modes} \cite{hawking2016soft} living on the black hole horizon. 

In what follows, we outline the construction of a subset of these soft excitations, generalizing the analysis of \cite{afshar2016soft}. The soft modes are defined as quantum excitations that carry the same energy as the ground state. In the fermionic extension of the theory, we explicitly present the corresponding energy operator and identify a large class of soft hair modes within this framework.

We then proceed to quantize the system and characterize the physical Hilbert space that accommodates these soft hair states. A systematic approach to quantization would involve formulating the reduced phase-space Lagrangian and applying canonical quantization. However, in this work we adopt a simpler route: we extend the known quantization procedure of the purely bosonic theory to the extended case, as detailed below.\footnote{See also \cite{chattopadhyay20222d, dutta2025bosonization}.}

We begin by introducing a convenient parameter that will play the role of the effective coupling constant,  
\begin{equation}
    \mathfrak{h} = \frac{4\pi}{\mathrm{k}}.
\end{equation}
The classical limit of the theory corresponds to $\mathfrak{h} \to 0$. 

Promoting the Poisson brackets in \eqref{eq:PoissonB} to quantum (anti-)commutators, we obtain the algebraic relations
\begin{equation}
\label{eq:PoissonBtilde}
\begin{split}
	[\,p_\pm (t,\theta),\, p_\pm (t,\theta') \,] &= \mp \, i \, \mathfrak{h} \, \delta'(\theta - \theta'), \\
	[\, p_\pm (t,\theta),\, \phi^\alpha_\pm (t,\theta') \,] &= \pm \, i \, \alpha \, \mathfrak{h} \, \phi_\pm^\alpha (t,\theta) \, \delta(\theta - \theta'), \\
	\{\phi^\alpha_\pm (t,\theta), \, \phi^\beta_\pm (t,\theta') \} &= \pm \, i \, \mathfrak{h} \, C^{\alpha \beta} \, \delta'(\theta - \theta') \, \pm \, i \tfrac{1}{2} \mathfrak{h} \, p_\pm(t,\theta) \, \delta^{\alpha + \beta, 0} \, \delta(\theta - \theta').
\end{split}
\end{equation}
These relations define the basic commutation and anti-commutation structure among the bosonic variables $p_\pm$ and their fermionic partners $\phi^\alpha_\pm$.

To analyze the quantum algebra, we now expand the fields in terms of Fourier modes:
\begin{align}
    p_\pm(t,\theta) &= \mathfrak{h} \sum_{m \, \in \, \mathbb{Z}} \, \alpha^\pm_{m}(t) \, e^{i m \theta }, \\
    \phi^\beta_\pm(t,\theta) & = \mathfrak{h} \sum_{r \, \in \, \mathbb{Z} + \frac{1}{2}} \, \varphi^\beta_{\pm ;\, r}(t) \, e^{i r \theta },
\end{align} 
where $\beta = \pm \tfrac{1}{2}$ labels the two fermionic components.  
As both $p_\pm$ and $\phi^{\pm \beta}_\pm$ are real fields, their modes satisfy the usual hermiticity conditions:
\begin{equation}
    \alpha^{\pm \, *}_{n} = \alpha^{\pm}_{-n}, \qquad \varphi^{\beta \, *}_{\pm;r} = \varphi^{\beta}_{\pm; -r}.
\end{equation}

The commutation and anti-commutation relations among the modes take the form
\begin{equation}
\label{eq:mode_algebra}
\begin{split}
    [\, \alpha^\pm_{m}, \, \alpha^\pm_{n} \,] & = \pm \, \frac{1}{2\pi \mathfrak{h}} \, m \, \delta_{m+n,0}, \\
    [\, \alpha^\pm_m ,\, \varphi^\beta_{\pm;\,r} \,] & = \pm \frac{i}{2\pi} \beta \,\varphi^\beta_{\pm;\, m + r}, \\
    \{ \varphi^\beta_{\pm; \,r},\, \varphi^\gamma_{\pm; \, s} \} & = \mp \, \frac{1}{2\pi \mathfrak{h}} \, r \, \delta_{r+s,0} \,C^{\beta \gamma} \,\pm\, \frac{i}{4\pi} \, \alpha^\pm_{r+s} \, \delta^{\beta + \gamma, 0},
    \end{split}
\end{equation} where the bosonic mode labels $m,n$ take integer values, while the fermionic mode labels $r,s$ are half-integers. All other (anti-)commutators vanish. 

To simplify the fermionic sector, it is convenient to introduce a new basis:
\begin{equation}
{\psi}^\dagger_\pm(t,\theta) \equiv  \phi^{ \frac{1}{2}}_\pm(t,\theta) + i \phi^{-\frac{1}{2}}_\pm(t,\theta), \qquad {\psi}_\pm(t,\theta) \equiv \phi^{ \frac{1}{2}}_\pm(t,\theta) - i \phi^{-\frac{1}{2}}_\pm(t,\theta).
\end{equation}
These combinations diagonalize the reality conditions and make the subsequent algebraic structure more transparent. Expanding these new fields in modes gives
\begin{equation}
    {\psi}^{\dagger}_\pm(t,\theta) = \mathfrak{h} \sum_{r \, \in \, \mathbb{Z} + \frac{1}{2}} {\psi}^\dagger_{\pm ; \,r}(t) \, e^{i r \theta }, \qquad {\psi}_\pm(t,\theta)= \mathfrak{h} \sum_{r \, \in \, \mathbb{Z} + \frac{1}{2}} {\psi}_{\pm ; \, r}(t) \, e^{-i r \theta },
\end{equation} 
which leads to the following relations between the old and new fermionic modes:
\begin{equation}
	\psi^\dagger_{\pm; r} = \varphi^\frac{1}{2}_{\pm; r} + i \varphi^{-\frac{1}{2}}_{\pm; r}, \qquad \psi_{\pm; r} = \varphi^\frac{1}{2}_{\pm; -r} - i \varphi^{-\frac{1}{2}}_{\pm; -r}.
\end{equation}

In terms of these variables, the mode algebra \eqref{eq:mode_algebra} simplifies to
\begin{equation}
\label{eq:algebrafinal}
\begin{split}
    [\, \alpha^\pm_m ,\, \psi^\dagger_{\pm; \,r} \,] & = \pm \frac{i}{4\pi} \, \psi_{\pm;\,-m-r}, \qquad [\, \alpha^\pm_m ,\, \psi_{\pm; \,r} \,] = \pm \frac{i}{4\pi} \, \psi^\dagger_{\pm;\,m-r},\\
    \{ \psi^\dagger_{\pm; \, r}, \, \psi^\dagger_{\pm; \, s}\} &=  \mp \, \frac{1}{2\pi} \, \alpha^\pm_{r+s}, \qquad \{ \psi_{\pm; \, r}, \, \psi_{\pm; \, s}\} =  \pm \, \frac{1}{2\pi} \, \alpha^\pm_{-r-s},\\
    \{ \psi^\dagger_{\pm; \, r}, \, \psi_{\pm; \, s}\} &= \mp \frac{i}{\pi \mathfrak{h}} \, r \, \delta_{r,s}.
\end{split}
\end{equation}
This defines a coupled algebra between the bosonic and fermionic oscillator modes.

The normal ordering prescription for fermions is chosen as
\begin{equation}
\label{eq:normalorder}
    :\psi^\dagger_{\pm; \,q} \, \psi_{\pm; \,r}: \ = 
    \begin{cases}
        ~~ \psi_{\pm; \, r} \, \psi^\dagger_{\pm; \, q}& r>0, \\
        - \psi^\dagger_{\pm; \,q} \, \psi_{\pm; \,r} & r<0 .
    \end{cases}
\end{equation}

We define the Fock vacuum $\ket{0}$ through the annihilation conditions
\begin{equation}
\label{eq:GSdef}
\begin{split}
    \alpha^\pm_n \ket{0} &= 0, \qquad n \geq 0,\\
    \psi^\dagger_{\pm; \,r} \ket{0} &= 0, \qquad r > 0, \\
   \psi_{\pm; \, r} \ket{0} &= 0, \qquad r < 0. 
\end{split}
\end{equation}
This ensures that creation and annihilation operators act consistently with the algebra \eqref{eq:algebrafinal}. 

We now turn to the construction of the Hamiltonians \eqref{Hamiltonian} governing the soft sector. To capture the essential features of the soft hair excitations, we consider the simplest possible form, with 
\begin{equation}
	g^{\pm}(p_{\pm}) = \vartheta_{\pm} \, p_{\pm},
\end{equation}
where $\vartheta_{\pm}$ are constants in the angular coordinate $\theta$. 

Under this assumption, the Hamiltonians can be written in terms of the mode operators as
\begin{align}
H^{\pm} &=  \vartheta_{\pm}(t) \, \alpha^{\pm}_0 (t) + \mathfrak{h}  \,c^{\pm}(t)\sum_{r\in \mathbb{Z}+\frac{1}{2}} : \varphi^\frac{1}{2}_{\pm; \, r}(t) \,\varphi^{-\frac{1}{2}}_{\pm; \, - r}(t): \\
&=  \vartheta_{\pm}(t) \, \alpha^{\pm}_0 (t)+ \frac{i \mathfrak{h}}{4} \,c^{\pm}(t) \sum_{r\in \mathbb{Z}+\frac{1}{2}}: \psi^\dagger_{\pm; \, r}(t) \, \psi_{\pm; \, r}(t) :
\end{align}
The first term represents the bosonic zero-mode contribution, while the second encodes the fermionic excitations around the ground state.

Finally, it is easy to verify that any vacuum descendant of the form
\begin{equation}
\prod_{n \in \mathbb{Z}_{\geq 0}} \, (\alpha^+_{-n})^{k^+_n} (\alpha^-_{-n})^{k^-_n}\ket{0}, \qquad k^\pm_n \in \mathbb{Z}_{\geq 0},
\end{equation}
remains a \emph{soft state} even in the presence of the fermionic contributions to the Hamiltonians. These states thus span the soft hair sector of the theory.

This result is more special than the purely bosonic case presented in \cite{afshar2016soft}, as this is necessarily true only for vacuum descendents for which the fermionic terms in the Hamiltonians do not have any energy contribution. In the bosonic case, one could replace $\ket{0}$ with any other state and draw similar conclusions.

\section{Discussion and outlook}
\label{sec:discussion}

In this work, we have constructed and analyzed two distinct fermionic extensions of the Kac--Moody boundary conditions for three-dimensional AdS$_3$ supergravity, formulated as a Chern--Simons gauge theory. Both constructions preserve the overall structure of the bosonic theory while incorporating fermionic degrees of freedom in a consistent supersymmetric framework. Despite their apparent similarity at the classical level, the two extensions exhibit crucial differences in their integrable structures and in their relation to known superconformal systems.

In the first construction, we implemented the supersymmetric generalization by employing the super--Miura transformation. This approach directly relates the boundary dynamics to the well-known super--Virasoro algebra. The super--Miura map provides a non-linear relation between the Kac--Moody currents and the super-Virasoro generators, allowing the corresponding hierarchy of conserved charges to be identified with the supersymmetric Gardner (or super--KdV) hierarchy. Consequently, the asymptotic symmetry algebra in this case is governed by the standard $\mathcal{N}=1$ super--Virasoro structure. The boundary dynamics can thus be interpreted as those of a supersymmetric integrable system, where the bosonic and fermionic sectors are coupled in a manner consistent with superconformal symmetry.

The second construction, in contrast, represents a genuinely new type of fermionic extension of the Kac--Moody boundary conditions. Here, the fermionic degrees of freedom enter the boundary phase space in a way that cannot be mapped to the super-Virasoro framework through a large gauge transformation. The resulting algebraic structure is not equivalent to the super-Virasoro algebra and instead realizes a different extension of the affine current algebra. This formulation admits a distinct set of Hamiltonians and conserved charges. Importantly, this new extension accommodates a broader class of boundary dynamics and may provide an alternative route to understanding fermionic boundary conditions beyond the conventional superconformal framework.

We also extend the droplet description of the collective field theory to the fermionic case, showing that fermionic excitations correspond to folded Fermi surfaces. The fermion bilinear $\mathrm{w}_\pm = \phi^{1/2}_\pm \phi^{-1/2}_\pm$ represents the first fold mode $w_{\pm 1}$, which evolves independently of higher moments. Thus, only $w_{\pm 1}$ remains dynamically relevant, while higher $w_{\pm n}$ modes are not generated.

We further carried out the quantisation of the extended Kac--Moody system by promoting the classical Poisson brackets to quantum (anti-)commutators, leading to a coupled boson--fermion oscillator algebra. The Fock vacuum and normal ordering prescriptions are introduced to define the quantum Hilbert space. The resulting Hamiltonians incorporate both bosonic zero modes and fermionic excitations. Remarkably, the spectrum contains only bosonic soft modes, indicating that the inclusion of fermionic terms does not generate additional fermionic soft excitations.

The analysis presented here opens several directions for future work. One natural extension is to investigate higher supersymmetric cases, such as $\mathcal{N}=(2,2)$ or $\mathcal{N}=(4,4)$ AdS$_3$ supergravity, and determine the corresponding super-integrable hierarchies that generalize the KdV-type structures mentioned here. Another promising avenue is to explore the holographic interpretation of the soft hair modes, including their role in microstate counting and their potential relation to near-horizon graded algebras. It would also be interesting to examine whether the newly discovered non--super-Virasoro extension admits a dual boundary description in terms of a deformed or non-conformal supersymmetric field theory.

Finally, the free Fermi description developed in this work provides a useful basis for constructing collective field formulations of the boundary dynamics. Such a reformulation may clarify the connection with super-Liouville or super-Toda systems and facilitate the study of quantized phase-space structures. Extending this framework to asymptotically flat or warped AdS$_3$ backgrounds could further illuminate the interplay between topology, supersymmetry, and the emergence of soft modes in three-dimensional quantum gravity.

We hope that the two fermionic extensions presented here, one directly linked to the super-Virasoro algebra and the other defining a genuinely new class of extended boundary dynamics, will serve as complementary frameworks for exploring the rich structure of AdS$_3$ supergravity and its quantum microstates.

\paragraph{Acknowledgment:} The work of NB is partially supported by SERB POWER fellowship SPG/2022/000370. VB would like to thank IISER Bhopal for hospitality where part of this work was done. We are indebted to people of India for their unconditional support toward the researches in basic science.

\appendix
\section{Conventions}
\label{appendix:basis}
The non-zero commutators of the 2d $\N=1$ superconformal algebra, also called the \textit{super-Virasoro algebra}, are given by,
\begin{equation}
\label{eq:N=1superconformal}
\begin{split}
    [L_{a}, L_{b}] &= (a - b) L_{a+b} + \frac{c}{12} \, a(a^2 - 1) \delta_{a + b, 0}, \\ 
    [L_{a}, Q_{\alpha}] &= \bigg(\frac{a}{2} - \alpha\bigg) Q_{a + \alpha}, \\
    \{Q_{\alpha}, Q_{\beta}\} &= -2L_{\alpha + \beta} - \frac{c}{3} \bigg(\alpha^2-\frac{1}{4} \bigg)\delta_{\alpha + \beta,0}.
\end{split}
\end{equation}

Here the latin indices $a,b,\ldots \in \zz$ and the greek indices $\alpha, \beta, \ldots \in \zz + \frac{1}{2}$ (Neveu-Schwarz). The global bulk subalgebra, consisting of the generators corresponding to $a,b = -1, 0, 1$ and $\alpha, \beta = \pm \frac{1}{2}$, is called $\osp(1|2,\mathbb{R})$. Explicitly,
\begin{equation}
\begin{split}
	[L_{a}, L_{b}] &= (a - b) L_{a+b}, \\ 
    [L_{a}, Q_{\alpha}] &= \bigg(\frac{a}{2} - \alpha\bigg) Q_{a + \alpha}, \\
    \{Q_{\alpha}, Q_{\beta}\} &= -2 L_{\alpha + \beta}.
\end{split}	
\end{equation}

When taking two copies of the algebras, we will sometimes put $\pm$ indices to denote the left and right sectors respectively. Commutators of generators from different sectors will always be zero.

In this basis for $\osp(1|2,\mathbb{R})$, we can choose the supersymmetric invariant non-degenerate bilinears as \cite{banerjee2017n, valcarcel2019new}
\begin{equation}
	\STr(L_a^\pm L_b^\pm) = \pm \frac{1}{2} \gamma_{ab}, \qquad \STr(Q_\alpha^\pm Q_\beta^\pm) = \pm 2 C_{\alpha \beta}.
\end{equation} where the tangent space metric is given by\begin{equation}
\label{tangentspacemetric}
	\gamma_{ab} = \begin{pmatrix}
		0 & 0 & -2 \\ 0 & 1 & 0 \\ -2 & 0 & 0
	\end{pmatrix},
\end{equation} and the charge conjugation matrix $C = i \sigma_2$ is given by,
\begin{equation}
	C^{\alpha \beta} = -C_{\alpha \beta} = \begin{pmatrix}
		0 & 1 \\ -1 & 0
	\end{pmatrix},
\end{equation} where the fermionic indices run over $\mp \frac{1}{2}$. Note that the bosonic sector of $\osp(1|2,\mathbb{R})$, generated by $L_a$ for $a \in \{-1, 0, 1\}$, is just $\mathfrak{sl}(2,\rr)$.

\bibliographystyle{hieeetr}
\bibliography{citations.bib}	
\end{document}